\documentclass{article}

 \pdfoutput=1

\usepackage{arxiv}

\usepackage[utf8]{inputenc} 
\usepackage[T1]{fontenc}    
\usepackage{hyperref}       
\usepackage{url}            
\usepackage{booktabs}       
\usepackage{amsfonts}       
\usepackage{nicefrac}       
\usepackage{microtype}      
\usepackage{lipsum}	      	
\usepackage{graphicx}
\usepackage{natbib}
\usepackage{doi}

\usepackage{amsmath}
\usepackage{amssymb}
\usepackage{amsthm}

\usepackage{bm}
\usepackage{algorithm} 
\usepackage{algpseudocode} 

\usepackage[dvipsnames]{xcolor}
\usepackage{floatpag}
\usepackage{caption}
\usepackage{subcaption}
\usepackage{adjustbox}
\usepackage{booktabs}
\usepackage{nicefrac}
\usepackage{soul}
\usepackage{mathtools}
\newcommand{\bs}[1]{\boldsymbol{#1}}

\DeclareMathOperator{\argmax}{argmax}
\DeclareMathOperator{\argmin}{argmin}

\DeclareMathOperator{\Corr}{Corr}
\DeclareMathOperator{\Cov}{Cov}
\DeclareMathOperator{\Var}{Var}

\DeclareMathOperator{\diag}{diag}

\newtheorem{theorem}{Theorem}[section]

\newtheorem{lemma}[theorem]{Lemma}
\newtheorem{definition}[theorem]{Definition}

\title{Efficient Computation of Sparse and Robust Maximum Association Estimators}

\date{}

\author{ \href{https://orcid.org/0000-0002-8492-7318}{\includegraphics[scale=0.06]{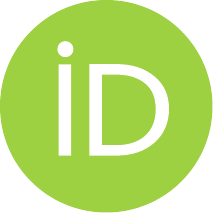}\hspace{1mm}Pia Pfeiffer \thanks{
    This work was funded by the Austrian COMET-Program (project InTribology1, no.~872176) via the Austrian Research Promotion Agency (FFG) and the federal states of Niederösterreich and Vorarlberg and was carried out at the Austrian Excellence Centre of Tribology (AC2T research GmbH) and the TU Wien. The authors acknowledge TU Wien Bibliothek for financial support through its Open Access Funding Programme.}\hspace{.2cm}} \\
        Institute of Statistics and \\ Mathematical Methods in Economics \\
        TU Wien \\
	\texttt{pia.pfeiffer@tuwien.ac.at} \\
	\And
	\href{https://orcid.org/0000-0002-2513-3788}{\includegraphics[scale=0.06]{orcid.pdf}\hspace{1mm}Andreas Alfons} \\
        Department of Econometrics \\
        Erasmus Universiteit Rotterdam \\
	\texttt{alfons@ese.eur.nl} \\
 \And
	\href{https://orcid.org/0000-0002-8014-4682}{\includegraphics[scale=0.06]{orcid.pdf}\hspace{1mm}Peter Filzmoser} \\
        Institute of Statistics and \\ Mathematical Methods in Economics \\
        TU Wien \\
	\texttt{peter.filzmoser@tuwien.ac.at} \\ }

\renewcommand{\shorttitle}{Sparse and Robust Maximum Association}

\hypersetup{
pdftitle={Efficient Computation of Sparse and Robust Maximum Association Estimators},
pdfsubject={stat.CO, stat.ML},
pdfauthor={Pia Pfeiffer, Andreas Alfons, Peter Filzmoser},
pdfkeywords={Biconvex optimization, Sparse robust canonical correlation, Robust estimation, Penalized canonical correlation},
}

\begin{document}
\maketitle

\begin{abstract}
Robust statistical estimators offer resilience against outliers but are often computationally challenging, particularly in high-dimensional sparse settings. Modern optimization techniques are utilized for robust sparse association estimators without imposing constraints on the covariance structure. The approach splits the problem into a robust estimation phase, followed by optimization of a decoupled, biconvex problem to derive the sparse canonical vectors. An augmented Lagrangian algorithm, combined with a modified adaptive gradient descent method, induces sparsity through simultaneous updates of both canonical vectors. Results demonstrate improved precision over existing methods, with high-dimensional empirical examples illustrating the effectiveness of this approach. The methodology can also be extended to other robust sparse estimators.
\end{abstract}


\section{Introduction}
\label{sec:intro}
With the availability of new measurement techniques, various different characteristics can be acquired from one and the same object. As an example from tribology, an engine oil can be investigated with respect to its chemical element composition, spectral information can be derived, or various properties concerning friction and wear of the oil can be measured, including image information of the degradation caused by the oil condition. Another example is biological data, specifically, the association between gene expressions and other variables, such as hepatic fatty acid concentrations related to a specific diet \citep[see, e.g.,][]{Martin2007}. The quantification of the relationships between different data sources can be very informative for a deeper understanding of already established mechanisms as well as for the generation of new hypotheses.

More formally, we are interested in the relationships between a $p$-dimen\-sional real-valued random vector $\bm x$ and a $q$-dimensional real-valued random vector $\bm y$. We consider the problem of obtaining coefficient vectors $\bm a$ and $\bm b$ such that the linear combinations $\bm a' \bm x$ and $\bm b' \bm y$ have maximum association, measured by an appropriate measure of association between univariate random variables.
A widely applied method for this task is canonical correlation analysis (CCA)~\citep[see, e.g.,][]{JohW07}.
The first canonical correlation coefficient $\rho_1$ and the first pair of canonical variables $(\bm a_1, \bm b_1)$ are defined via the maximization of the correlation coefficient between the two linear combinations \citep[see, e.g.,][]{JohW07}, that is,
\begin{align}
    \rho_1 &= \max_{\substack{\bm a, \bm b \\ \|\bm a\|= \|\bm b\|=1} } \Corr(\bm a' \bm x, \bm b' \bm y)  \label{eq:cancor_def}, \\ 
    (\bm a_1, \bm b_1)&= \argmax_{\|\bm a\|=1,\|\bm b\|=1} \Corr(\bm a' \bm x, \bm b' \bm y). \label{eq:CCA}
\end{align}
The $k$-th canonical correlation coefficient $\rho_k$ and the respective pair of canonical variables $(\bm a_k, \bm b_k)$ are obtained similarly as in (\ref{eq:cancor_def})--(\ref{eq:CCA}), but under the additional constraint that they are uncorrelated with the previous $k-1$ directions, for $k\in \{2,\ldots, \min(p,q)\}$.
Expression (\ref{eq:cancor_def}) can be written in terms of the covariance:
\begin{align}
    \max_{\substack{\bm a, \bm b \\ \|\bm a\|= \|\bm b\|=1} } \Corr(\bm a' \bm x, \bm b' \bm y)
            &= \max_{\substack{\bm a, \bm b \\ \|\bm a\|= \|\bm b\|=1} } \frac{\Cov(\bm a' \bm x, \bm b' \bm y)}{\sqrt{\Var(\bm a' \bm x)}\sqrt{\Var(\bm b' \bm y)}} \nonumber \\
            &= \max_{\substack{\bm a, \bm b \\ \|\bm a\|= \|\bm b\|=1} }\frac{\bm a' \bm \Sigma_{xy}\bm b}{\sqrt{\bm a'\bm \Sigma_{xx}\bm a}\sqrt{\bm b'\bm \Sigma_{yy}\bm b}}, \label{eq:cov_form}
\end{align}
where $\bm \Sigma_{xx} = \Cov(\bm x)$, $\bm \Sigma_{yy} = \Cov(\bm y)$ and $\bm \Sigma_{xy} = \Cov(\bm x, \bm y)$.
The analytical solution is given by the eigenvectors and eigenvalues of a combination of (inverse) covariance matrices: 
$\rho_i^2$ are eigenvalues of $\bm \Sigma_{xx}^{-1}\bm \Sigma_{xy}\bm \Sigma_{yy}^{-1}\bm \Sigma_{yx}$ with normed eigenvectors $\bm a_i$, and $\rho_i^2$ are also eigenvalues of $\bm \Sigma_{yy}^{-1}\bm \Sigma_{yx}\bm \Sigma_{xx}^{-1}\bm \Sigma_{xy}$ with normed eigenvectors $\bm b_i$, for $i=1,\ldots ,\min(p,q)$ \citep[see, e.g.,][]{JohW07}.

Classically, the involved covariance matrices are estimated by the sample covariances, and this corresponds to maximizing the Pearson correlation coefficient as measure of association. However, these estimators are sensitive to outlying observations, and the solution is not well-defined in the high-dimensional setting, when more variables than observations are available.

There are several approaches in the literature to derive a robust solution. For the \emph{plug-in approach}, the sample covariance is replaced by a robust estimator of the joint covariance of ${\bm x}$ and ${\bm y}$. \citet{Croux2002} propose to use the minimum covariance determinant (MCD) estimator \citep{Rousseeuw1984, Rousseeuw1985}, and they derive influence functions for the canonical correlations and vectors based on this plug-in estimator, revealing their robustness properties.
For a broader class of affine equivariant scatter and shape matrices, influence functions and limiting distributions of canonical correlations and vectors have been studied by \citet{Taskinen2006}. \citet{Langworthy2020} present theoretical results about using the transformed Kendall correlation, which is more robust under violation of the normality assumption, for the estimation of a scatter matrix.

Another approach is to generalize (\ref{eq:cancor_def}) to a wider class of association estimators. \citet{Alfons2016} modify the optimization problem (\ref{eq:cancor_def}) in a robust way, and also consider rank-correlation measures such as the Spearman rank correlation. In that way, the search for linear relationships, as done with the Pearson correlation, is extended to looking for non-linear relationships. 
Results concerning Fisher consistency and the influence function underline the good theoretical properties of the corresponding robust maximum association measures, which represent the strongest association between linear combinations of two sets of random variables. The optimization is done using a  grid algorithm \citep{Alfons2016b} which, however, has its limitations concerning the dimensionality $p$ and $q$ of the two random variables.

The high-dimensional case, when more variables than observations are present, is another scenario where the sample covariance matrix performs poorly. 
This can be addressed by regularizing the covariance matrix as in the penalized matrix decomposition (PMD) method of \citet{Witten2009}, where the relationship between the singular value decomposition (SVD) and the Frobenius norm is exploited and optimization is done via a soft-thresholded power method. \citet{Chen2013} develop a canonical pair model and a sparse power algorithm combined with iterative thresholding is applied to estimate the precision matrices.
The \emph{alternating regression approach} \citep{Waaijenborg2008, Wilms2015a} avoids the computation of covariance matrices and considers problem~(\ref{eq:cancor_def}) from a predictive point of view. \citet{Wilms2015a} derive sparse directions by applying sparsity-inducing regression estimators like the least absolute shrinkage and selection operator (LASSO) \citep{Tibshirani1996} or its robustification, sparse least trimmed squares (sparseLTS), introduced by \citet{Alfons2013}.
\citet{Gu2020} combine the alternating regression approach with an alternating direction method of multipliers (ADMM) algorithm for an $L1$ penalized setting, and \citet{Gao2017} propose an adaptive estimation method in the high-dimensional setting, which is numerically solved by an ADMM algorithm.
\citet{Shu2020} describe a CCA method suitable for high-dimensional data based on methods identifying common and distinctive components such as joint and individual variation explained (JIVE) or simultaneous component analysis with rotation to common and distinctive components (DISCO-SCA).

\citet{Mai2019} propose an alternative problem formulation based on the least squares loss and additional scaling terms that is solved via an iterative least squares algorithm. However, this method also needs the covariance matrix as an input, and the least squares loss is not robust. For a robust variant based on their approach, both the covariance matrix would need to be estimated in a robust way, and the squared loss function would need to be substituted with a robust alternative, leading to robust alternating regressions as proposed by \citet{Wilms2015b}, however, with the added problem of estimating the covariance robustly. Similar concerns hold for a variant using trace LASSO regularization, as proposed by \citet{Deng2020}. The objective function is again based on a least squares loss function and, therefore, non-robust. In addition, a non-robust covariance estimator is applied for the constraints. \citet{Tuzhilina2023} show a ``kernel'' trick for regularized CCA with a ridge penalty, but their approach is not suitable for penalties that induce sparsity in the canonical vectors.

To the best of our knowledge, the only robust \textit{and} sparse method that does not require the repeated computation and inversion of high-dimensional covariance matrices is based on robust alternating regressions \citep{Wilms2015b}. For higher-order associations, however, there is no efficient implementation available. 
Several authors propose to use \emph{deflated data matrices} for computing higher-order correlations \citep{Alfons2016, Wilms2015b}. However, this approach requires solving several regression problems and can potentially destroy sparsity. \citet{Wilms2015b} address this by applying a sparsity-inducing regression estimator. 

The optimization problems (\ref{eq:cancor_def})--(\ref{eq:cov_form}) based on robust correlation or estimators of the covariance matrix lead to highly non-convex objective functions. To obtain a problem formulation that is easier to optimize, the robust estimation and the optimization are \emph{decoupled}: In the first step, the covariance matrix is estimated robustly. This robust estimate of the covariance matrix is then plugged into the subsequent problem formulation with sparsity constraints on the canonical vectors, yielding a biconvex problem.  

Note that our aim is to provide easily interpretable canonical vectors, hence we introduce sparsity only in the canonical vectors but not in the covariance matrix.
\citet{Witten2009} suggest a similar formulation of the optimization problem for the non-robust case, and an iterative method is presented. Our method, however, also considers the denominator in (\ref{eq:cov_form}), and offers flexibility in the choice of sparsity constraints as well as for the estimator of the covariance matrix. 
Since rank-based estimators of the covariance matrix will be considered as well, we will use the terminology ``(robust) association measure'' instead of ``canonical correlation coefficient'', and simply ``linear combinations'' instead of ``canonical vectors'' in the following.

The remainder of the paper is organized as follows: First, the reformulation of the problem is detailed, and an appropriate algorithm for its numerical solution is introduced. Then, the results of a simulation study are presented to illustrate the suitability of our approach for a high-dimensional setting with outliers and to compare its performance to existing approaches. We conclude with an outlook on other common statistical tasks that can be solved by applying the algorithm in a similar way. 

\section{Robust and sparse maximum association}

\subsection{Formulation as a constrained optimization problem}

The optimization problems stated in Section~\ref{sec:intro} can also be formulated as constrained optimization problem \citep[see, e.g.,][]{Anderson1958}.
This problem formulation has the advantage that the conditions for uncorrelatedness for directions of higher order and sparsity-inducing penalty terms  for the coefficient vectors of the linear combination can be stated directly and added as constraints. Starting from expression~(\ref{eq:cov_form}), $\bm \Sigma_{xx}, \bm \Sigma_{yy},$ and $\bm \Sigma_{xy}$ are substituted with suitable estimators for the covariance, denoted by $\bm C_{xx}, \bm C_{yy},$ and $\bm C_{xy}$.
Then, the first order maximum association coefficient $\rho_1$ and the corresponding vectors $(\bm a_1, \bm b_1)$ can be obtained as a solution to the following optimization problem:
\begin{align}
    \min_{\substack{ \bm a \in \mathbb{R}^p, \bm b \in \mathbb{R}^q}} - F(\bm a, \bm b) \label{eq:CCA_opt}
\end{align}
with $F: \mathbb{R}^p \times \mathbb{R}^q \rightarrow \mathbb{R}:~F(\bm a, \bm b) = \bm a' \bm C_{xy} \bm b$ under the constraints
\begin{align}
    \bm a' \bm C_{xx} \bm a &= 1 \label{eq:cancor_const_a}, \\
    \bm b' \bm C_{yy} \bm b &= 1 \label{eq:cancor_const_b}.
\end{align} 
This problem formulation avoids the repeated evaluation of the correlation measure that is needed for a projection-pursuit approach as suggested by \citet{Alfons2016}. The covariance needs to be estimated only once to act as input for the optimization problem, and is then fixed for the optimization process.
For higher-order coefficients $\rho_k$ and vectors $(\bm a_k, \bm b_k)$, $k\in \{2, \ldots ,\min(p,q)\}$, constraints for uncorrelatedness with the lower-order directions are needed:
\begin{align}
    \bm a_k' \bm C_{xx} \bm a_i &= 0, \qquad i = 1, \ldots , k-1, \label{eq:cancor_const_lowa}\\
    \bm b_k' \bm C_{yy} \bm b_i &= 0, \qquad i = 1, \ldots , k-1. \label{eq:cancor_const_lowb}
\end{align}
Especially for the high-dimensional setting, where $p$ and/or $q$ are big, it can be desirable to set some coefficients to zero in the vectors for the linear combinations. Thus, penalty terms can be added as further constraints in the form of
\begin{align}
    P_{a_k}(\bm a_k) &\leq c_{a_k}, \label{eq:cancor_const_pena} \\
    P_{b_k}(\bm b_k) &\leq c_{b_k}, \label{eq:cancor_const_penb}
\end{align}
where $c_{a_k}$ and $c_{b_k}$ denote positive constants. Here, the penalty terms (\ref{eq:cancor_const_pena})--(\ref{eq:cancor_const_penb}) are taken as elastic net penalties
\begin{align}
    P_{a_k}(\bm u) &= \alpha_{a_k} \|\bm u\|_1 + (1-\alpha_{a_k})\bm \|\bm u\|_2^2,
    \label{eq:cancor_eneta}\\
    P_{b_k}(\bm u) &= \alpha_{b_k} \|\bm u\|_1 + (1-\alpha_{b_k})\bm \|\bm u\|_2^2,
    \label{eq:cancor_enetb}
\end{align}
but other (convex) penalties are also applicable.

\citet{Witten2009} also suggest formulating CCA as an optimization problem and derive the canonical directions via an iterative power method. Our approach is more general in that (i) there are no additional assumptions imposed on the covariance matrix, and (ii) the penalty function can be adapted for each order and can also differ for $\bm a$ and $\bm b$. 

\subsection{Robust estimation of the covariance matrix}

The choice of a suitable estimator of the covariance matrix is crucial for obtaining robust estimators of the canonical vectors \citep{Alfons2016}. The robustness of the estimator of the covariance matrix and its stability in the high-dimensional case will influence the respective properties of the resulting coefficients and vectors \citep{Taskinen2006}.
In this work, the focus is on the following estimators: For a base result, the sample covariance matrix is used to estimate $\bm \Sigma_{xx}, \bm \Sigma_{yy},$ and $\bm \Sigma_{xy}$, which corresponds to using the Pearson correlation coefficient as measure of association. To achieve robustness and to allow for the high-dimensional case, the minimum regularized covariance determinant (MRCD) \citep{Boudt2020} and orthogonalized Gnanadesikan-Kettenring (OGK) \citep{Maronna2002} estimators are used to estimate the joint covariance matrix of $\bm x$ and $\bm y$, which is afterward decomposed into the matrices $\bm C_{xx}, \bm C_{yy},$ and $\bm C_{xy}$. The MRCD estimator is based on minimizing the determinant of a \emph{regularized} covariance matrix over all possible subsets of a given size $h \leq n$, and it can also be seen as a robust version of the Ledoit-Wolf estimator \citep{Ledoit2004}. The OGK estimator relies on applying the identity $\Cov(x, y) = (\sigma(x+y)^2 - \sigma(x-y)^2)/4$, where $\sigma$ is the standard deviation and $x$ and $y$ denotes a pair of random variables. This identity is applied for the pairwise combinations of the components in the joint vector of $\bm x$ and $\bm y$, by using a robust scale estimator. The final robust estimator of the covariance matrix is obtained after an orthogonalization step. Note that this result is not necessarily positive definite and eigenvalue correction to obtain a positive definite estimator of the covariance matrix is applied. For both the MRCD and the OGK estimator, the implementations in the package \texttt{rrcov} \citep{R_rrcov} for the statistical computing environment R \citep{R_language} are used. The OGK estimator is thereby applied with the default settings (using the initial covariance as proposed by \citet{Gnanadesikan1972} and the $\tau$ scale \citep{Yohai1988} for univariate location and dispersion). For MRCD, the size of the $h$ subset, controlled by the parameter $\alpha$, is set to 75\% of the number of observations.

As an alternative, pairwise correlation estimators based on Spearman's rank and Kendall's \textit{tau} are also investigated. They can easily be computed in the high-dimensional case as well and have desirable robustness properties \citep{Croux2010, Alfons2016}. Denote $\bm R$ as the resulting correlation matrix of the joint vector of $\bm x=(x_1,\ldots ,x_p)'$ and $\bm y=(y_1,\ldots ,y_q)'$, and $\bm D = \diag(\sigma(x_1), \ldots, \sigma(x_p), \sigma(y_1), \ldots, \sigma(y_q))$, where $\sigma$ corresponds to a (robust) scale estimate. Then the joint covariance is obtained as $\bm D \bm R \bm D$. For $\sigma$ we used the median absolute deviation (MAD). Note that
for consistency at the normal distribution, it is necessary to apply the transformation $s_{ij}=\frac{6}{\pi} \arcsin \left(\frac{r_{ij}^S}{2} \right)$ to the raw Spearman's rank correlation coefficient $r_{ij}^S$, and the transformation $\tau_{ij} = \frac{2}{\pi} \arcsin (r_{ij}^K)$ to the raw Kendall's \textit{tau} coefficient $r_{ij}^K$, where the indices $i$ and $j$ refer to a pair of univariate variables. 
As \citet{Langworthy2020} point out, a potential issue with those covariance matrices based on pairwise estimation is that they are not necessarily positive definite. Various methods have been proposed to adjust the estimated covariance matrix so that it is positive definite. \citet{Rousseeuw1993}, for example, discuss transformations based on shrinkage and eigenvalues. \citet{Higham2002} presents a method to find the nearest correlation matrix in the Frobenius norm. This algorithm is implemented as the function \texttt{nearPD()} in the R package \texttt{Matrix} \citep{R_Matrix}. While the positive definiteness of the estimator of the covariance matrix is necessary for the existence of a solution, the presented algorithm does not rely on this property for the computation of the maximum association and corresponding linear combinations. However, the results may not be reliable and in our implementation of the proposed algorithm in the R package \texttt{RobSparseMVA} (see Section~\ref{sec:algorithm} for more information), it is possible to include a check for positive definiteness and apply the \texttt{nearPD()} transformation in case the assumption is violated before starting the optimization algorithm.
While the Gaussian rank correlation \citep{Iman1982} avoids complications with positive definiteness, it compromises the robustness: The asymptotic breakdown point for the Gaussian rank correlation is $\varepsilon^*(\hat{\rho}_G) = 12.4\%$ \citep{Boudt2012}, compared to $\varepsilon^*(\hat{\rho}_S) = 20.6\%$ for the Spearman correlation, and $\varepsilon^*(\hat{\rho}_K) = 29.3\%$ for Kendall's $\tau$ \citep{Grize1978, Davies2005}. As we are considering applications that need a high level of robustness, in the following, we consider only the more robust rank correlation measures.

\subsection{Lagrangian formulation and properties}

The conditions (\ref{eq:cancor_const_a})--(\ref{eq:cancor_const_b}) in the constrained optimization problem (\ref{eq:CCA_opt})--(\ref{eq:cancor_const_penb}) are not convex. However, they can be modified to be convex by replacing the equality constraint with an inequality constraint:
\begin{align}
    \bm a' \bm C_{xx} \bm a &\leq 1 \tag{\ref{eq:cancor_const_a}a} \label{eq:cancor_const_a_ie}\\
    \bm b' \bm C_{yy} \bm b &\leq 1 \tag{\ref{eq:cancor_const_b}a} \label{eq:cancor_const_b_ie}.
\end{align} 
The modified optimization problem is now biconvex (that is, convex in $\bm a$ if $\bm b$ is fixed and vice versa) and has the same solution as the original problem under the condition that the constants $c_{a_k}$ and $c_{b_k}$ are chosen such that $\bm a' \bm C_{xx} \bm a \geq 1$ and $\bm b' \bm C_{yy} \bm b \geq 1$ hold (see, e.g., \citealp{Boyd2005}, as cited by \citealp{Witten2009}).
This can be accomplished for any solution by ensuring that the pair $(\bm a, \bm b)$ minimizing \eqref{eq:CCA_opt} subject to \eqref{eq:cancor_const_lowa}--\eqref{eq:cancor_const_penb} is scaled by $\sqrt{\bm a' \bm C_{xx} \bm a}$ and $\sqrt{\bm b' \bm C_{yy} \bm b}$, respectively.

The Lagrangian function related to the optimization problem (\ref{eq:CCA_opt})--(\ref{eq:cancor_const_penb}) is given by
\begin{align}
\mathcal{L}(\bm a, \bm b, \bs{\lambda}) = -\bm a' \bm C_{xy}\bm b + \bs{\mu}' \cdot G(\bm a, \bm b) + \bs{\lambda}' \cdot H(\bm a, \bm b) ,\label{eq:Lagrange} 
\end{align}
with the equality constraints given by
\begin{align}
H: \mathbb{R}^p \times \mathbb{R}^q \rightarrow \mathbb{R}^{2k -2}: 
    H(\bm a, \bm b) = \left [\begin{matrix} 
          \bm a' \bm C_{xx} \bm a_{1:(k-1)} \\
          \bm b' \bm C_{yy} \bm b_{1:(k-1)} 
       \end{matrix} \right ], \label{eq:constraint_H}
\end{align}
and the inequality constraints summarized by
\begin{align}
G: \mathbb{R}^p \times \mathbb{R}^q \rightarrow \mathbb{R}^{4}: 
    G(\bm a, \bm b) = \left [\begin{matrix} 
          \bm a' \bm C_{xx} \bm a - 1\\
          \bm b' \bm C_{yy} \bm b - 1 \\
          P_1(\bm a) - c_a \\
          P_2(\bm b) - c_b 
       \end{matrix} \right ],  \label{eq:constraint_G}
\end{align}
and the Karush-Kuhn-Tucker (KKT) multipliers denoted by $\bm \lambda$ (dimension $2k-2$) and $\bm \mu$ (4-dimensional), respectively.

The optimization problem attains a minimum over the set of points that satisfy the constraints \eqref{eq:cancor_const_a}--\eqref{eq:cancor_const_penb}. In addition, the Mangasarian-Fromovitz constraint qualification
 (MFCQ) \citep{Mangasarian1967} holds for all $(\bm a, \bm b) \neq \bm 0$, resulting in all minima being regular points. These are standard conditions \citep[see, e.g.,][]{Ruszczynski2006}, and the proofs for the given optimization problem are provided in Appendix~\ref{app:lemmas}. A minimizer of the original optimization problem \eqref{eq:CCA_opt}--\eqref{eq:cancor_const_penb} therefore has to satisfy the KKT conditions \citep{Boyd2005}.

Another interesting property is the inherent regularization of the covariance matrix when a penalty is imposed on the canonical vectors.
For obtaining the first order association coefficients and vectors, the Lagrange multipliers are $\bs{\lambda}=(\lambda_1,\ldots ,\lambda_4)'$, 
and by setting the derivative of $\mathcal{L}$ to $\bm 0$, we obtain
\begin{align}
    -\bm C_{xy}\bm b + \lambda_1 \bm C_{xx} \bm a + \lambda_3 \frac{\partial}{\partial \bm a}P_{\bm a_1} (\bm a) &= \bm 0, \label{eq:rel1}\\
    -\bm C_{yx}\bm a + \lambda_2 \bm C_{yy} \bm b + \lambda_4 \frac{\partial}{\partial \bm b}P_{\bm b_1} (\bm b) &= \bm 0. \label{eq:rel2}
\end{align}
When $P_{\bm a_k}$ and $P_{\bm b_k}$ are given as ridge penalties, the derivatives can be written as
\begin{align}
    \frac{\partial}{\partial \bm u} P(\bm u) = 2\bm I \bm u.
\end{align}
Substitution in Equations \eqref{eq:rel1} and \eqref{eq:rel2} followed by applying an inverse transformation yields
\begin{align}
\left [ \bm C_{xx} + \frac{\lambda_3}{\lambda_1} \bm I \right ]^{-1} \bm C_{xy} \left [ \bm C_{yy} + \frac{\lambda_4}{\lambda_2} \bm I \right ]^{-1} \bm C_{yx} \bm a &= \lambda_1 \lambda_2 \bm a, \label{eq:reg_cxx}\\
\left [ \bm C_{yy} + \frac{\lambda_4}{\lambda_2} \bm I \right ]^{-1}\bm C_{yx}  \left [ \bm C_{xx} + \frac{\lambda_3}{\lambda_1} \bm I \right ]^{-1}\bm C_{xy} \bm b &= \lambda_1 \lambda_2 \bm b. \label{eq:reg_cyy}
\end{align}
It can be seen that this formulation corresponds to a regularization of the estimators of the covariance matrix $\bm C_{xx}$ and $\bm C_{yy}$. 
However, when the regularization is applied explicitly to the vectors (and only implicitly to the covariance matrix), solving the resulting equations is not trivial: When a more general penalty than the ridge penalty is used, the estimating equations would involve finding the inverse of a non-linear operator, which is the case even for an $L1$- or elastic net penalty. 
As an alternative, we propose an iterative algorithm that is based on a combination of the method of multipliers (MM) and a state-of-the-art gradient-based algorithm for the minimization of the resulting sequence of unconstrained problems.

\section{Algorithm} \label{sec:algorithm}

Using the \emph{augmented Lagrangian} or \emph{method of multipliers (MM)} \citep[see, e.g.,][]{Boyd2005}, the problem can be rewritten as a minimization problem in an unconstrained form. The MM-algorithm has been studied extensively by \citet{Bertsekas1982}, and the ADMM variation has been brought back more recently due to its potential for distributed computing \citep{Boyd2011}. The main idea of the MM approach is to convert the constrained optimization problem to a series of unconstrained problems.
The augmented Lagrangian function is given by 
\begin{align}
\mathcal{L}_c(\bm a, \bm b, \bm \mu, \bm \lambda) = & -F(\bm a, \bm b) + \bm \mu' \cdot G(\bm a, \bm b) +  \bm \lambda' \cdot H(\bm a, \bm b) \\ 
&+ \frac{c}{2} (\| H(\bm a, \bm b)\|_2^2 + G(\bm a, \bm b))\|_2^2) , \label{eq:augLagrange} 
\end{align}
where $F$ denotes the primal objective. The equality constraints are given by \eqref{eq:constraint_H} and the inequality constraints are given by \eqref{eq:constraint_G}. 

The corresponding multipliers are denoted by $\bm \mu$ and $\bm \lambda$, and the strength of the regularization term for the equality constraints is given by $c \in \mathbb{R}$.
Given the current iterates $(\bm a^t, \bm b^t)$, $\bm \mu^t$ and $\bm \lambda^t$, the updated $(\bm a^{t+1}, \bm b^{t+1})$ are derived via minimization of $\mathcal{L}_c(\bm a^t, \bm b^t, \bm \mu^t, \bm \lambda^t)$. Then, the dual variables $\bm \mu$ and $\bm \lambda$ are updated. The complete algorithm is given in Algorithm~\ref{alg:robsparse_cca}, Appendix~\ref{app:alg}.

 As the solution to the minimization problem 
 \begin{align}
     (\bm a_k^{t+1}, \bm b_k^{t+1}) \leftarrow \argmin ~\mathcal{L}_c(\bm a_k^t, \bm b_k^t; \bm \mu^t,  \bm \lambda^t)
 \end{align} in line \ref{alg:optimize_args} of Algorithm~\ref{alg:robsparse_cca}, Appendix~\ref{app:alg}, cannot be derived analytically in the general case, the minimization is done by adaptive gradient descent as introduced by \citet{Kingma2014} and refined by \citet{Reddi2018}. We implemented the minimization step based on the AMSGrad optimizer and implemented in the R package \texttt{torch} \citep{R_torch}. The complete algorithm can be found in Algorithm~\ref{alg:amsgrad}, Appendix~\ref{app:alg}. In the following, we highlight important modifications to the original algorithm. Most notably, the update of $\bm a$ and $\bm b$ is done in one step. To accomplish this, we use the gradient of the partitioned vector $\bm z = (\bm a, \bm b)'$ together with the embeddings $\bm a = \bm E_a'\bm z$ and $\bm b = \bm E_b' \bm z$, where 
 \begin{align}
    [\bm E_a | \bm E_b] =  
        \begin{bmatrix}
            \bm I_p & \bm 0 \\
            \bm 0 & \bm I_q
        \end{bmatrix}.
    \end{align}
 The constraints are already partitioned, and the primal objective can be formulated in terms of $\bm z$, namely as
 \begin{align}
     F(\bm z) = \bm z' \bm E_a \bm C_{xy} \bm E_b' \bm z,
 \end{align}
 yielding a loss function depending on $\bm z$. 

 Other gradient-based optimizers could also be applied. Methods using an adaptive learning rate and momentum such as AMSGrad or Adam are preferred choices, as they are capable of escaping local optima and are less sensitive to the initial choice of the learning rate $\alpha_0$. All constraints are subdifferentiable (i.e., for all points in the domain of $\mathcal{L}_c$, at least one subgradient exists), and the subgradient update as implemented in \texttt{torch} can be executed. In addition, a thresholding step is included in the algorithm (lines \ref{alg:thresh_a}--\ref{alg:thresh_b}) to get true sparsity, which is not possible from the subgradient update alone. Note that projected gradient methods are not easily combined with adaptive learning algorithms, as the computed momentum and scaling would need to be adjusted for the projected gradient. Thresholding the resulting linear combinations provides an effective alternative. Thresholding is done using the moving average of the last $M$ step sizes; 
 in practice we were successful with setting $M=10$.
 Depending on the current value of $H$, the regularization parameter $c$ is updated in lines \ref{alg:reg_if}--\ref{alg:reg} of Algorithm \ref{alg:robsparse_cca}. The constants 0.25 and 10 in lines \ref{alg:reg_if}--\ref{alg:reg} were already proposed by \citet{Bertsekas1982} and work well in our simulations.

 \subsection{Convergence to stationary point}

Biconvex optimization problems are commonly treated in an alternating manner, for the problem (\ref{eq:CCA_opt})--(\ref{eq:cancor_const_penb}) that would suggest updating $\bm a$ while fixing $\bm b$ and vice versa (this course of action would correspond to the ADMM algorithm).
Even though the partial problems are convex, in general, there is no guarantee that the partial optimums derived from an alternating search are even stationary points of the original problem \citep{Gorski2007}.
Therefore, instead of alternating the updates of $\bm a$ and $\bm b$, we propose to perform the update at the same time with a gradient-descent-type algorithm.  This way, the algorithm converges towards a stationary point of the Lagrange function $\mathcal{L}_0$, which can be seen as follows: Via gradient descent, a stationary point of $\mathcal{L}_c(\bm a_k^t, \bm b_k^t; \bm \mu^t, \bm \lambda^t)$, denoted by $(\bm a_k^{t+1}, \bm b_k^{t+1})$, is identified. This point fulfills
\begin{align}
     0 \in &~\partial \mathcal{L}_c(\bm a_k^{t+1}, \bm b_k^{t+1}; \bm \mu^t, \bm \lambda^t) = \\
      = &-\partial F(\bm a_k^{t+1}, \bm b_k^{t+1}) \\
     &+ \bm \mu ^{t'}\partial G(\bm a_k^{t+1}, \bm b_k^{t+1}) + cG(\bm a_k^{t+1}, \bm b_k^{t+1})\partial G(\bm a_k^{t+1}, \bm b_k^{t+1}) \\
     &+ \bm \lambda ^{t'}\partial H(\bm a_k^{t+1}, \bm b_k^{t+1}) + cH(\bm a_k^{t+1}, \bm b_k^{t+1})\partial H(\bm a_k^{t+1}, \bm b_k^{t+1}).
\end{align}
With the updates $\bm \mu^{t+1} = \bm \mu^t + cG(\bm a_k^{t+1}, \bm b_k^{t+1})$ and $\bm \lambda^{t+1} = \bm \lambda^t + cH(\bm a_k^{t+1}, \bm b_k^{t+1})$, it follows that $0 \in \partial \mathcal{L}_0(\bm a_k^{t+1}, \bm b_k^{t+1}; \bm \mu^{t+1}, \bm \lambda^{t+1})$.
Here, $\partial \mathcal{L}_c$ refers to the subgradient of the loss function $\mathcal{L}_c$. When the loss function is differentiable, the subgradient is identical to the gradient $\nabla \mathcal{L}_c$. In the subdifferentiable setting, a stationary point fulfills $0 \in \partial \mathcal{L}_c$, as the subgradient is defined as a set of vectors (see Appendix~\ref{app:prelim} for more details on the definition of the subgradient).

\begin{figure}
     \centering
     \begin{subfigure}{1\textwidth}
         \includegraphics{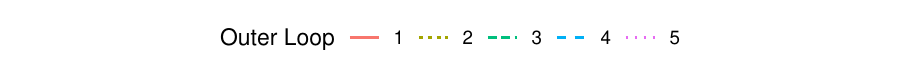}
     \end{subfigure}
     \begin{subfigure}{0.495\textwidth}
         \includegraphics{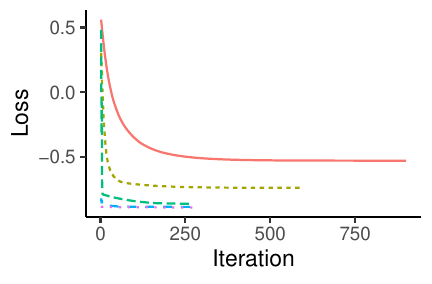}
     \end{subfigure}
     \begin{subfigure}{0.495\textwidth}
         \includegraphics{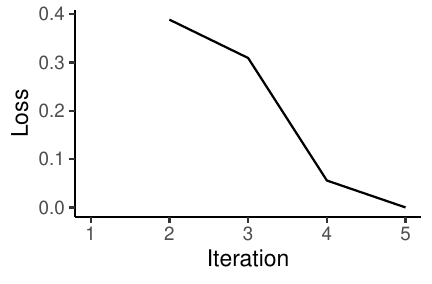}
     \end{subfigure}
     \caption{Visualization of the convergence of the inner loop (left) and the outer loop (right), where the inner loop corresponds to the minimization of $\mathcal{L}_c$ and the outer loop the multiplier update. The results from the Figures show the loss values for the first order associations and were computed on a low-dimensional (see Section~\ref{sec:simdesign}) and clean example with fixed $L1$-penalties $\| \bm a\|_1 \leq 1.4$ and $\| \bm b\|_1 \leq 1.4$.}
     \label{fig:convergence_viz}
 \end{figure}

 In the optimization problem derived from classical CCA, only the canonical vectors (of all orders) satisfy the stationarity condition. Theoretically, the algorithm could end up in a stationary point corresponding to the maximum association and the respective linear combinations of a different order. However, our simulations indicate that this is not a problem in practice, as the applied variant of gradient descent can escape local optima. It is advisable to observe the optimization path and, if need be, adjust the parameters of the algorithm. In case the directions are derived in the wrong order, they can easily be sorted by applying a robust measure of association. Visualizations of the loss functions for both the inner and outer optimization loop of an example setting are given in Figure~\ref{fig:convergence_viz}. 

\subsection{Complexity}\label{sec:complexity}

By means of the big $O$ notation, the computational complexity of the different approaches can be compared. For the proposed algorithm, the complexity consists of the complexity of the gradient-descent algorithm, given by $O(tp)$ for an assumed number of $t$ iterations and $p$ variables ($p > q$), and the covariance estimation, depending on the estimator. For PMD, the complexity depends on the eigenvalue computation ($O(p^2)$ for the power iteration) plus the covariance estimation. For SRAR, the complexity depends on the type of robust regression estimator ($O(\nicefrac{n^p}{h})$). Especially for data with a high number of variables $p$ and $q$, gradient descent has advantages compared to a robust regression approach. Note that to exploit this, it is necessary to combine the algorithm with a fast robust covariance estimator, like covariance matrices based on rank-correlation matrices. In Appendix~\ref{app:complexity}, we provide a more extensive discussion of computational complexity for the different approaches and a number of choices of covariance estimator. The findings are summarized in Table~\ref{tab:complexity}.

\subsection{Hyperparameter optimization}
 Another important aspect of the algorithm is the choice of the hyperparameters. The mixing parameters $\alpha_{a_k}$ and $\alpha_{b_k}$ of the elastic net penalties in (\ref{eq:cancor_eneta}) and (\ref{eq:cancor_enetb}) are often set in advance by the user, but the sparsity parameters $c_{a_k}$ and $c_{b_k}$ have to be determined in a data-driven manner.
 This selection has to be done in two dimensions (for $\bm a$ and $\bm b$) and for each order of vectors and associations to be determined.
 Grid search in combination with cross-validation quickly becomes infeasible if the search space becomes larger, especially in more than one dimension. An alternative is \emph{Bayesian optimization} of the given hyperparameters.
 An introduction to Bayesian optimization for hyperparameter optimization can be found, for example, in \citet{Frazier2018}. The benefit of using Bayesian optimization instead of grid search is that the information from previous function evaluations can be used to determine the best next point to execute the function. This way, a much bigger search space can be covered, and, in addition, it is less likely to miss a good parameter configuration due to the size of the grid. 
 In Appendix~\ref{app:b_opt}, more details on the Bayesian optimization algorithm, as well as choices for the acquisition score functions, are described.

\subsection{Initialization}

For the presented algorithm, suitable starting values $\bm a_k^0$ and $\bm b_k^0$ for the linear combinations $\bm a_k$ and $\bm b_k$, respectively, and for the associated Lagrange multiplier $\bm \lambda_k$ are needed. 
For the elements of $\bm a_1^0$ and $\bm b_1^0$, we use the average contribution of the respective row or column in $\bm C_{xy}=[c(x_i,y_j)]$ as a starting value, 
\begin{align}
      a^0_{1_i} = \frac{1}{q} \sum_{j=1}^q c(x_i,y_j)
      \quad \mbox{ for } \quad i=1,\ldots ,p\ ,\\
      b^0_{1_j} = \frac{1}{p} \sum_{i=1}^p c(x_i,y_j) 
      \quad \mbox{ for } \quad j=1,\ldots ,q\ ,
\end{align}
and for the Langrange multipliers, the contraints are evaluated at the starting points,
\begin{align}
      \bm \lambda_k^0 = H(\bm a^0_k, \bm b^0_k) .
\end{align}
 
If a non-robust estimator of the covariance matrix is used, the starting values may already be influenced by outlying observations. A more detailed investigation of what happens when the number of outlying observations is increased is given in Section~\ref{sec:increasing_cont}.

For the computation of directions of higher order, the concept of "deflated" data matrices is often described in the literature \citep[e.g.][]{Branco2005}. 
However, this step can be detrimental to the sparsity in the higher-order vectors.
In our approach, constraints for uncorrelatedness to lower-order directions are added to the model. Higher-order directions need to satisfy Equations (\ref{eq:cancor_const_lowa}) and (\ref{eq:cancor_const_lowb}), respectively. Basically, this means that $\bm a_k$ is in the left null space of $\bm C_{xx}\bm a^{(i:k-1)}$ and $\bm b_k$ is in the left null space of $\bm C_{yy}\bm b^{(i:k-1)}$. These affine constraints preserve the biconvexity and suggest the following variation for determining the starting values for higher-order linear combinations: The orthogonal complements $\bm A^\perp_k:= \{\bm a: \bm a'\bm C_{xx}\bm a^{(i:k-1)} = 0\}$ and $\bm B^\perp_k:= \{\bm b: \bm b'\bm C_{yy}\bm b^{(i:k-1)} = 0 \}$ are computed, then the starting values $\bm a_k^0$ and $\bm b_k^0$ are chosen as the orthogonal projections of $\bm a_1^0$ and $\bm b_1^0$ on $\bm A^\perp_k$ and $\bm B^\perp_k$, respectively. 

In our simulations, both the naive approach and this "orthogonal" initialization for the higher-order linear combinations are compared. 
 
\section{Simulation study} 
\label{sec:simulation_study}

A simulation study was conducted to compare the performance of the proposed method using different (robust) estimators of the covariance matrix. The comparison is also done with other approaches already mentioned in Section~\ref{sec:intro}, namely PMD by \citet{Witten2009}, SCCA by \citet{Mai2019}, and SRAR by \citet{Wilms2015a}. For PMD the R package \texttt{PMA} \citep{R_PMA} was used, for SCCA (sparse CCA) the code available from \url{https://academic.oup.com/biometrics/article/75/3/734/7537599#supplementary-data}, and for SRAR the code available from \url{https://sites.google.com/view/iwilms/software}. Our algorithm is implemented in the R package \texttt{RobSparseMVA} and available online from \url{https://github.com/piapfeiffer/RobSparseMVA}.

A good sparse and robust method should be efficient when the number of variables grows, avoid misidentifying important variables, and should attain these properties in the presence of outliers in the data \citep[see, e.g.,][]{Zou2006, Todorov2013}.
In order to check those requirements, the following \emph{performance measures} are used.
For measuring accuracy, the angle $\theta_a = \arccos\left(\frac{\bm a' \hat{\bm a}}{\|\bm a\|\cdot\|\hat{\bm a}\|}\right)$ between the true and estimated canonical variables is computed. Note that only the results for one of the linear combinations are presented here as the results for the other are qualitatively similar.
The true-positive rate (TPR), corresponding to the rate of correctly identified non-zero components, together with 
the true-negative rate (TNR), or the rate of correctly identified zero components, measure whether non-zero variables are identified correctly. For studying the scalability, the runtime for a growing number of variables is measured. 

For the computation of above performance measures, the true linear combinations $\bm a$ and $\bm b$ have to be computed: They can be derived from the true covariance matrices $\bm \Sigma_{xx}$, $\bm \Sigma_{xy}$, and $\bm \Sigma_{yy}$ as eigenvectors of $\bm \Sigma_{xx}^{-1} \bm \Sigma_{xy} \bm \Sigma_{yy}^{-1} \bm \Sigma_{yx}$ and $\bm \Sigma_{yy}^{-1} \bm \Sigma_{yx} \bm \Sigma_{xx}^{-1} \bm \Sigma_{xy}$, respectively, see Section~\ref{sec:intro}. 

\subsection{Simulation design}
\label{sec:simdesign}

Different simulation settings and contamination scenarios \citep[similar to][]{Wilms2015b} are considered. 
Clean data are generated from a multivariate normal distribution: $(\bm x,\bm y)' \sim \mathcal{N}_{p+q}(\bm 0, \bm \Sigma)$. For the contaminated scenario, $c_r$\% contamination is generated from a multivariate normal distribution with a mean shift: $(\bm x,\bm y)' \sim \mathcal{N}_{p+q}(c_s \cdot \bm{1}, \bm \Sigma)$, where $c_s$ denotes the contamination strength.
To simulate a heavy-tailed distribution, data are generated from a multivariate t-distribution: $(\bm x,\bm y)' \sim t_3(\bm 0, \bm \Sigma)$.
The joint covariance matrix
$\bm \Sigma = \left[ 
\begin{array}{cc}
\bm \Sigma_{xx} & \bm \Sigma_{xy} \\
\bm \Sigma_{xy}' & \bm \Sigma_{yy}
\end{array} \right]$ is given according to the 
following simulation settings:
\begin{enumerate}
    \item Low-dimensional, order 2: $p = q = 10, n = 100$ observations
    \begin{align*}
             \bm \Sigma_{xx} &= \bm I_{10} \\
             \bm \Sigma_{yy} &= \bm I_{10} \\
             \bm \Sigma_{xy} &= \left[ {\begin{array}{ccc}
             0.9 &  0 & \bm 0_{1\times 8} \\
              0 & 0.7 & \bm 0_{1\times 8}\\
              \bm 0_{8\times 1} & \bm 0_{8\times 1} & \bm 0_{8\times 8} \end{array} } \right]
             \end{align*}
    The true associations in this setting are $\rho_1 = 0.9$ and $\rho_2 = 0.7$, and the true linear combinations are $\bm a_1 = \bm b_1 = (1, \bm 0_{1\times9})'$ and $\bm a_2 = \bm b_2 = (0, 1, \bm 0_{1\times 8})'$.
    
    \item High-dimensional, order 2: $p = q = 100, n = 50$ observations
        \begin{align*}
             \bm \Sigma_{xx} &= \left[ {\begin{array}{ccc}
             \bm{S^1}_{10\times 10} &  \bm 0_{10\times 10} & \bm 0_{10\times 80} \\
              \bm 0_{10\times 10} & \bm{S^2}_{10 \times 10} & \bm 0_{10\times 80}\\
              \bm 0_{80\times 10} & \bm 0_{80\times 10} & \bm I_{80\times 80} \end{array} } \right] \\
             \bm \Sigma_{yy} &= \bm \Sigma_{xx} \\
             \bm \Sigma_{xy} &= \left[ {\begin{array}{ccc}
             \bm{0.9}_{10\times 10} &  \bm 0_{10\times 10} & \bm 0_{10\times 80} \\
              \bm 0_{10\times 10} & \bm{0.5}_{10 \times 10} & \bm 0_{10\times 80}\\
              \bm 0_{80\times 10} & \bm 0_{80\times 10} & \bm 0_{80\times 80} \end{array} } \right]
        \end{align*}
        where $\bm{S^1}_{ij}= 1$ if $i=j$ and $\bm{S^1}_{ij}= 0.9$ for $i\neq j$ and $\bm{S^2}_{ij}= 1$ if $i=j$ and $\bm{S^2}_{ij}= 0.7$ for $i\neq j$.
        The true associations in this setting are $\rho_1 = 0.989$ and $\rho_2 = 0.685$, and the true linear combinations are $\bm a_1 = \bm b_1 = (\bm {0.105}_{1\times10}, \bm 0_{1\times90})'$ and $\bm a_2 = \bm b_2 = (\bm 0_{1\times10}, \bm {0.117}_{1\times10}, \bm 0_{1\times80})'$.
\end{enumerate}

\subsection{Simulation results}

\subsubsection{Precision of the algorithm} \label{sec:precision}

For evaluating the precision of the algorithm, we avoid estimating the covariance matrix but plug in the true 
covariance matrix $\bm \Sigma$ into the algorithm. We also compare the precision when the theoretically optimal sparsity parameters are provided (which can be derived from the true linear combinations $(\bm a, \bm b)$ as $c_a = \| \bm a\|_1$ and $c_b =\| \bm b\|_1$) and when the sparsity parameters are estimated via Bayesian optimization. The results of these (deterministic) computations are presented in Table~\ref{tab:precision}.
For the first-order association measure, the algorithm always converges to the true solution. For higher-order association measures, it can be seen that the sparsity parameters need to be chosen with more care and that the orthogonal start is necessary for the high-dimensional setting. Furthermore, it can be concluded that starting with an orthogonal projection for the second-order directions is beneficial. Therefore, the results of all subsequent simulations are shown for this initialization.

\begin{table}[h!]
\caption{Performance measures given the true covariance. The true association measures for the low-dimensional and high-dimensional setting are $\rho_1 = 0.9$ and $\rho_2 = 0.7$, and $\rho_1 = 0.989$ and $\rho_2 = 0.685$, respectively.}
\label{tab:precision}
\resizebox{\textwidth}{!}{
\begin{tabular}{llllcccc}\toprule
\textbf{Setting} & \textbf{Sparsity} & \textbf{Initialization} & \textbf{Order} & \textbf{Angle $\theta_a$} & \textbf{TPR} & \textbf{TNR} & \textbf{Association}\\
\midrule
Low-dimensional & known & naive & 1 & 0 & 1 & 1 & 0.9\\
 &  & naive  & 2 & 0 & 1 & 0.89 & 0.7\\
 &  & orthogonal & 2 & 0 & 1 & 1 & 0.7\\
 & estimated & naive & 1 & 0 & 1 & 1 & 0.9\\
 &  & naive  & 2 & 0 & 1 & 0.89 & 0.7\\
 &  &orthogonal & 2 & 0 & 1 & 1 & 0.7\\
High-dimensional & known &naive & 1 & 0 & 1 & 1 & 0.989\\
 &  &naive & 2 & 1.57 & 0 & 0.89 & 0.989\\
 &  &orthogonal & 2 & 0.23 & 1 & 1 & 0.683\\
 & estimated &naive & 1 & 0 & 1 & 1 & 0.989\\
 &  &naive & 2 &  1.57 & 0 & 0.89 & 0.989\\
 &  &orthogonal & 2 & 0 & 1 & 1 & 0.685\\
 \bottomrule
\end{tabular}}
\end{table}

\subsubsection{Comparison to other methods}
For the given scenarios and settings, the proposed method using different estimators of the covariance matrix is compared to  SRAR by \citet{Wilms2015b}, sparse CCA via  PMD by \citet{Witten2009}, and SCCA by \citet{Mai2019}, in terms of estimation accuracy as measured by the angle $\theta_a$, and sparsity control as measured by the TPR and TNR. 
For our algorithm, we used the orthogonal initialization to compute the second-order association. 
The naive initialization leads to worse results (not shown here).

\begin{figure}[!htp]
\thisfloatpagestyle{empty}
\centering
\begin{adjustbox}{minipage=\textwidth, scale = 0.95}
\begin{subfigure}{1\textwidth}
    \includegraphics[width = 1\textwidth]{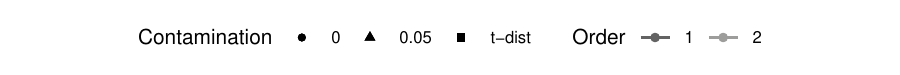}
\end{subfigure}

\begin{subfigure}{0.495\textwidth}
    \includegraphics[width = 1\textwidth]{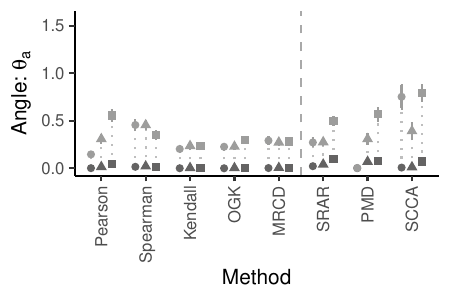}
\end{subfigure}
\begin{subfigure}{0.495\textwidth}
    \includegraphics[width = 1\textwidth]{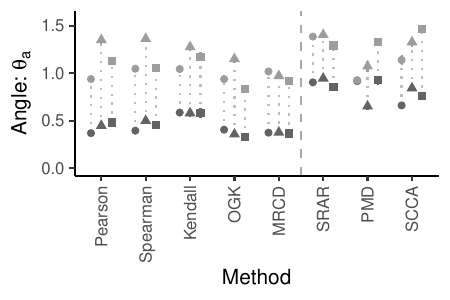}
\end{subfigure}

\begin{subfigure}{0.49\textwidth}
    \includegraphics[width = 1\textwidth]{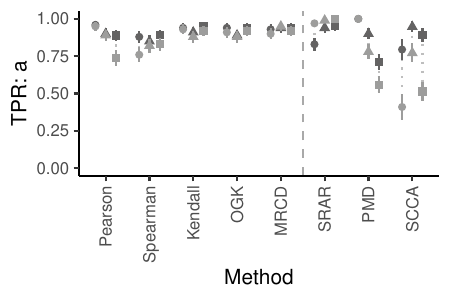}
\end{subfigure}
\begin{subfigure}{0.49\textwidth}
    \includegraphics[width = 1\textwidth]{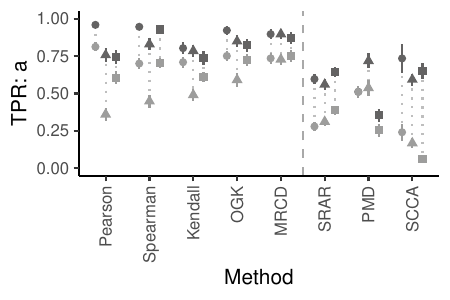}
\end{subfigure}

\begin{subfigure}{0.49\textwidth}
    \includegraphics[width = 1\textwidth]{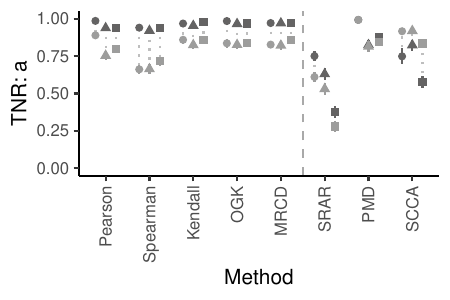}
\end{subfigure}
\begin{subfigure}{0.49\textwidth}
    \includegraphics[width = 1\textwidth]{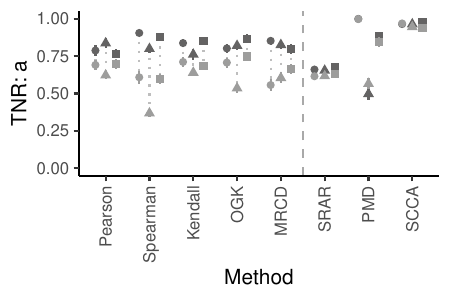}
\end{subfigure}
    \begin{subfigure}{0.49\textwidth}
    \includegraphics[width = 1\textwidth]{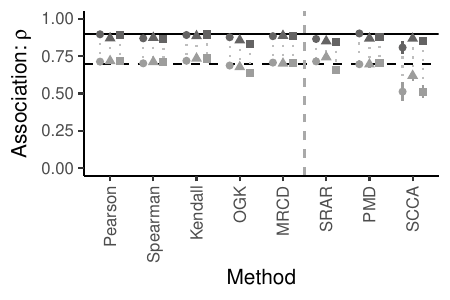}
\end{subfigure}
\begin{subfigure}{0.49\textwidth}
    \includegraphics[width = 1\textwidth]{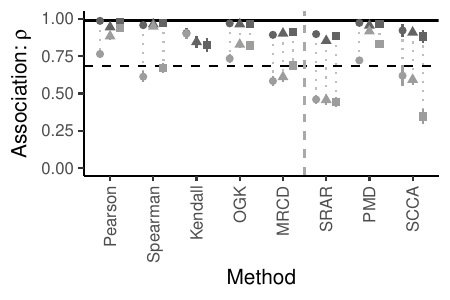}
\end{subfigure}
\caption{The proposed method ccaMM using various covariance matrix estimators (left-hand side of the vertical dashed line) is compared to SRAR, PMD, and SCCA in terms of estimation accuracy (angle $\theta_a$) and sparsity (TPR, TNR), as well as the estimated association measure ($\rho$).}
    \label{fig:comparison_methods}
    \end{adjustbox}
\end{figure}

The results over 100 repetitions for estimators of the covariance matrix used in our algorithm (left of the dashed line) with SRAR \citep{Wilms2015b}, PMD \citep{Witten2009}, and SCCA \citep{Mai2019} are summarized in Figure~\ref{fig:comparison_methods}. The left column presents the results for the low-dimensional setting, the
right column for the high-dimensional setting. 
The different plot symbols encode different contamination scenarios; black presents first-order results, and gray second-order results. 
Shown are the mean values over 100 repetitions for the metrics, together with error bars representing the standard error range.
Note that the standard errors are small so that the error bars are often barely visible.
We present the results for uncontaminated data,
for contamination of 5\% of the observations with contamination strength $c_s=2$, and for data generated from 
a multivariate $t_3$ distribution, see Section~\ref{sec:simdesign}.

The results for the low-dimensional setting are shown on the left-hand side of Figure~\ref{fig:comparison_methods}. The angle, TPR, and TNR are only presented for the estimated canonical variables $\bm a_1$ and $\bm a_2$. For the non-contaminated data, the performance across all methods is similar. Estimating the second-order component leads to slightly worse results - with the exception of PMD, where the results are highly precise. The behavior under contamination and heavy tails is still comparable to the non-contaminated case; only SRAR and PMD have problems identifying the correct sparsity.
For PMD, this is especially apparent in the figures depicting the TPR and TNR: Both for the first-order and second-order linear combinations, the TNR is 1, but the TPR is only around 0.5, indicating that the resulting linear combinations are too sparse. Similar conclusions can be drawn for SCCA, especially for the second-order components, for which this method performs worse across all performance metrics.

For the high-dimensional setting (right-hand side of Figure~\ref{fig:comparison_methods}), 
more differences can be observed: When the data are uncontaminated, our algorithm based on the Pearson correlation is highly accurate. This can be in part attributed to the regularization effect on the estimators of the covariance matrix described in Equations (\ref{eq:reg_cxx})--(\ref{eq:reg_cyy}). 
While the performance for the high-dimensional setting is overall worse, and all methods suffer from a decrease in accuracy for the second-order canonical vectors, it can be observed that for the robust OGK and MRCD estimators, the accuracy level for the second-order component is the same as it already is for the first-order component for SRAR (Figure~\ref{fig:comparison_methods}: top-right).
Our algorithm with robust estimators of the covariance matrix shows superior performance for the TPR. Especially interesting is the good result for the covariance based on Spearman's rank and Kendall's $tau$ in the scenario using the t-distribution. This finding coincides with the work presented in \citet{Langworthy2020}. 
The TNR is good for all variants of our algorithm and for PMD in the clean setting. SRAR performs worse, and similarly to the accuracy, the TNR for the second-order components estimated using our algorithm is on the same level as the TNR for SRAR in the first-order components.
Difficulties with accuracy and estimating the correct sparsity are also reflected in the resulting association 
measure. Especially SRAR and SCCA result in a worse estimate of the association measure.

\subsubsection{Increasing contamination}\label{sec:increasing_cont}
\begin{figure}[!htp]
\centering
\begin{adjustbox}{minipage=\textwidth, scale = 0.95}
\begin{subfigure}{1\textwidth}
    \includegraphics[width = 1\textwidth]{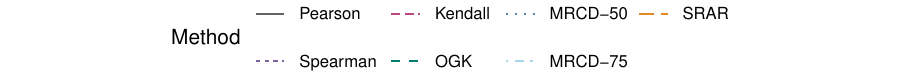}
\end{subfigure}

\begin{subfigure}{0.49\textwidth}
    \includegraphics[width = 1\textwidth]{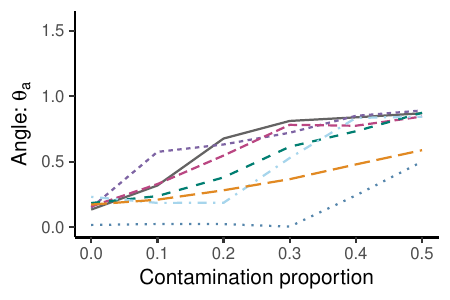}

\end{subfigure}
\begin{subfigure}{0.49\textwidth}
    \includegraphics[width = 1\textwidth]{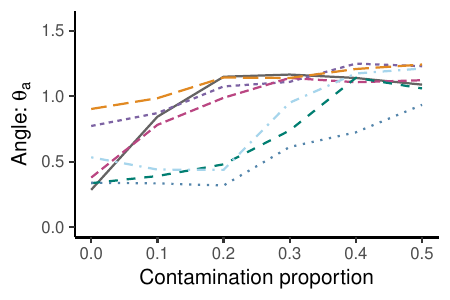}
\end{subfigure}

\begin{subfigure}{0.49\textwidth}
    \includegraphics[width = 1\textwidth]{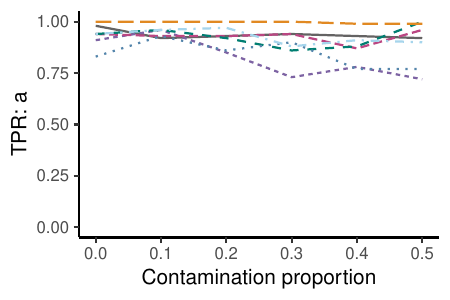}
\end{subfigure}
\begin{subfigure}{0.49\textwidth}
    \includegraphics[width = 1\textwidth]{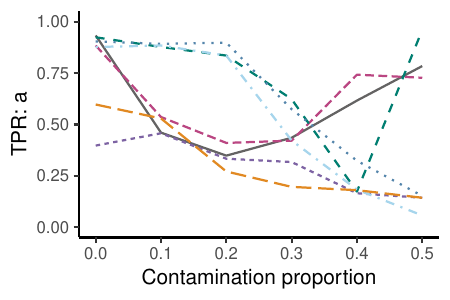}
\end{subfigure}

\begin{subfigure}{0.49\textwidth}
    \includegraphics[width = 1\textwidth]{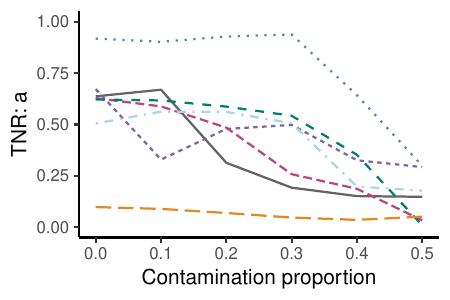}
\end{subfigure}
\begin{subfigure}{0.49\textwidth}
    \includegraphics[width = 1\textwidth]{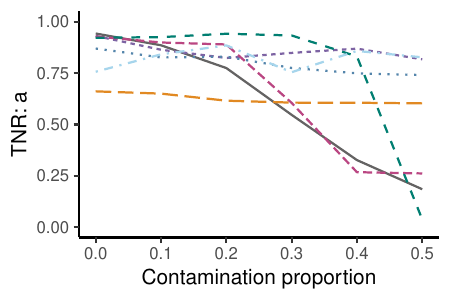}
\end{subfigure}

\begin{subfigure}{0.49\textwidth}
    \includegraphics[width = 1\textwidth]{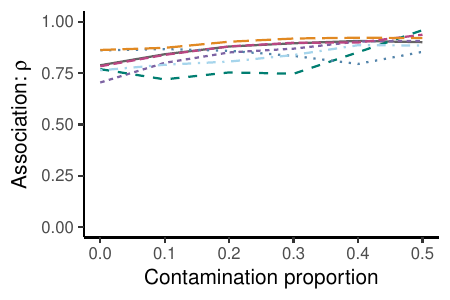}
\end{subfigure}
\begin{subfigure}{0.49\textwidth}
    \includegraphics[width = 1\textwidth]{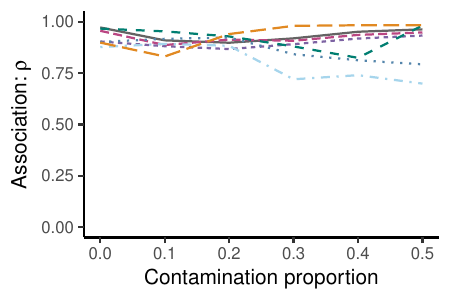}
\end{subfigure}
\caption{The proposed method ccaMM using various covariance matrix estimators is compared to SRAR and PMD in terms of estimation accuracy (angle $\theta_a$) and sparsity (TPR, TNR), as well as the estimated association measure ($\rho$) when the contamination ratio is increased.}
    \label{fig:increasing_cont}
    \end{adjustbox}
\end{figure}

For both the low-dimensional and high-dimensional settings described in Section~\ref{sec:simdesign}, the contamination proportion was increased from $cp = 0\%$ to $cp = 50\%$, and again, the same performance measures were evaluated. The results averaged over 50 repetitions comparing the different robust estimators are shown in Figure~\ref{fig:increasing_cont}. On the left side, the performance metrics for the low-dimensional setting are given, on the right side, the results for the high-dimensional setting are shown. The different line types correspond to the performance metrics for our method using different estimators for the covariance matrix (Pearson, Spearman, Kendall, OGK, MRCD) and the sparse and robust alternating regressions (SRAR) technique on the other hand. The results for the penalized matrix decomposition (PMD) and for sparse CCA (SCCA) are omitted, as these methods are already affected by a small proportion of outlying observations.
While the metrics for accuracy show in the low-dimensional setting an advantage for the alternating regressions approach, the other metrics are comparable across the methods. For the high-dimensional setting, the results are more distinguished: The proposed method outperforms the SRAR approach, while the robustness against an increased contamination proportion depends on the estimator of the covariance matrix. In these plots, it is clearly visible that the default setting of MRCD, i.e., $\alpha = 0.75$, denoted by MRCD-75 in the plot, performs much worse as the contamination proportion is increased to values greater than $cp = 25\%$, because the breakdown point
of the MRCD estimator is then just 25\%, see \citet{PuchF24}. When $\alpha = 0.5$ (denoted by MRCD-50), 
the breakdown point is approximately 50\%, and
the proposed approach in combination with MRCD outperforms the other methods. The OGK estimator also exhibits interesting behavior: The TPR metric decreases up to a contamination proportion of $40\%$, then it jumps to $1$. 
The reason can be observed when analyzing the TNR metric: After a contamination proportion of $40\%$ is reached, all components are identified to be non-zero. 
Another interesting observation is that while for all estimators the performance metrics decline, the predicted association measure still seems to be reliable.
The contamination strength (mean shift) was also varied between $c_s = 2$ and $c_s = 10$, but was not found to have an effect on the performance of the different estimators. Therefore, these results are omitted here.

\subsubsection{Runtime}
For evaluating the runtime, the first-order canonical directions and association were computed using our algorithm and SRAR for increasing dimension $q = 50, \ldots, 10000$ while keeping the dimensionality of the other side, $p = 10$, and the number of samples, $n=100$, fixed. 
The joint covariance matrix 
$\bm \Sigma = \left[ 
\begin{array}{cc}
\bm \Sigma_{xx} & \bm \Sigma_{xy} \\
\bm \Sigma_{xy}' & \bm \Sigma_{yy}
\end{array} \right]$
is given according to the settings:

 \begin{align*}
             \bm \Sigma_{xx} &= \left[ 
             \bm{S^1}_{10\times 10}   \right] \\
             \bm \Sigma_{yy} &= \left[ {\begin{array}{cc}
             \bm{S^1}_{10\times 10} & \bm 0_{10\times (q-10)} \\
              \bm 0_{(q-10)\times 10}  & \bm I_{(q-10)\times (q-10)} \end{array} } \right] \\
             \bm \Sigma_{xy} &= \left[ {\begin{array}{cc}
             \bm{0.8}_{10\times 10}  & \bm 0_{10\times (q-10)} \\
              \bm 0_{(q-10)\times 10} & \bm 0_{(q-10)\times (q-10)} \end{array} } \right]
        \end{align*}
        where $\bm{S^1}_{ij}= 1$ if $i=j$ and $\bm{S^1}_{ij}= 0.8$ for $i\neq j$.
    
\begin{figure}[!ht]
    \centering
    \includegraphics[width = 1\textwidth]{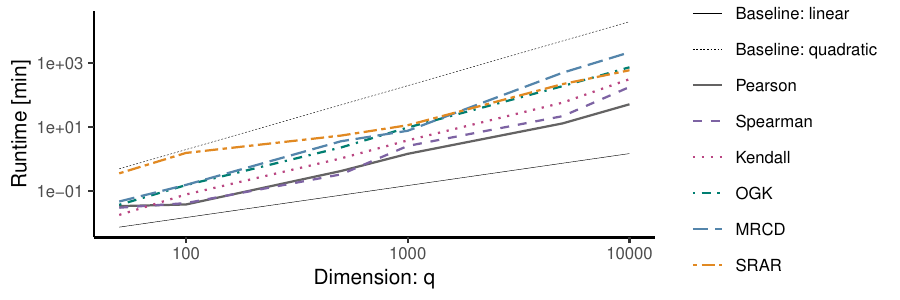}
    \caption{Log-log plot of CPU runtime in minutes versus dimensionality $q$ of second variable.}
    \label{fig:runtime}
\end{figure}

To remove the effect of the hyperparameter search, fixed optimal sparsity parameters were used for all methods. The results are averaged over 10 replications and presented in Figure~\ref{fig:runtime}. The increasing dimension $q$ is shown on the horizontal axis on a log scale, and the CPU runtime of the algorithm is shown on the vertical axis in minutes. It can be observed that the presented algorithm based on adaptive gradient descent has a comparable dependence of runtime to problem size to the only other robust and sparse alternative. The runtime depends on the type of association estimator, which aligns with the results on complexity from Section~\ref{sec:complexity}. In combination with rank-based estimators for the covariance, the proposed algorithm presents a more time-efficient alternative to alternating regressions.

In summary, the simulation results demonstrate the comparable or better performance of our proposed method, depending on the contamination scenario. Especially for the high-dimensional scenario, our method in combination with an appropriate estimator for the covariance matrix performs well, also for higher-order directions.  When runtime is of concern, the method based on Spearman or Kendall correlation should be considered, as these measures have better robustness properties than Pearson correlation, while still being efficient to compute. In general, the OGK estimator seems to be a sensible choice, leading to good performance metrics in both low-dimensional and high-dimensional settings and for an increased contamination ratio.

\section{Examples}
In this section, the proposed method will be applied to two data sets from different fields: biology and tribology. For both, the number of observations is significantly lower than the number of variables, and the flexibility of choosing the sparsity via the elastic-net penalty for each side is desirable.

In order to compare the performance of classical and robust estimation, we compare the out-of-sample performance of the robust and non-robust estimators by randomly splitting the data into a training and test set and computing the out-of-sample residual score 
\begin{equation}
\label{eq:residscore}
    r = \frac{1}{n_{\text{test}}}\frac{1}{n_{\text{rep}}} \sum_{j=1}^{n_{\text{test}}} \sum_{i=1}^{n_{\text{rep}}} \|\bm a_{\text{train}_i}'\bm x_{j} - \bm b_{\text{train}_i}'\bm y_{j}\|^2 ,
\end{equation}
where $\bm a_{\text{train}_i}$ and $\bm b_{\text{train}_i}$ are the first order linear combinations that are estimated on the 
$i$-th training set, $n_{\text{test}}$ is the number of test set observations, and $\bm x_{j}$ and $\bm y_{j}$ are observations
from the corresponding test sets. The random training and test splits are repeated $n_{\text{rep}}$ times.
Since outliers in the test sets could contaminate this residual score measure, we also use a trimmed version, by trimming the largest 10\% of all the squared test residuals and dividing by the corresponding number of observations. We also report the (in-sample) association measure, averaged over $n_{\text{rep}}$ random training and test splits.

\subsection{Application to the \texttt{nutrimouse} dataset}

The \texttt{nutrimouse} dataset is publicly available via the \texttt{CCA} package in R \citep{R_CCA} and has been discussed in the related literature, for example, by \citet{Wilms2015b}. It contains $n = 40$ observations of $p=120$ gene expressions and $q = 21$ concentrations of fatty acids. \citet{Martin2007} provide a detailed description of the dataset, and investigate the influence of a certain diet on numerous gene expressions in mice. In this setting, the goal is to identify a set of genes that has a large association with a  set of lipids \citep[cf.][]{Wilms2015b}. The two datasets were robustly centered and scaled with median and MAD before continuing with the analysis.
We compare the results of the sparse CCA method by \citet{Witten2009} with the proposed method by computing the residual score $r$, see Equation~(\ref{eq:residscore}) in a leave-one-out cross-validation (CV) setting, i.e., the $i$-th training set consists of all observations except $i$, and the $i$-th test set only contains observation $i$.

\begin{figure}[!ht]
\centering
        \includegraphics[width = 0.5\textwidth]{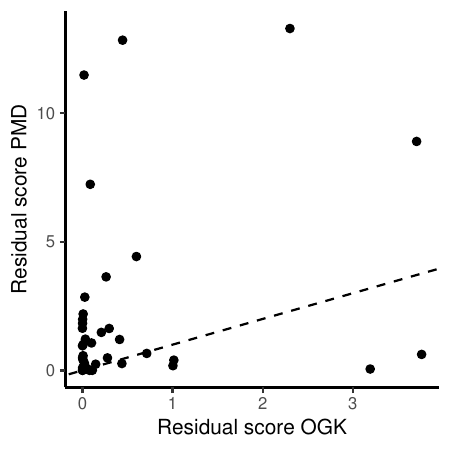}
    \caption{Scatterplot of leave-one-out cross-validation (CV) scores for PMD and the proposed method using the OGK estimator for the \texttt{nutrimouse} dataset. Almost all points are above the 45$^{\circ}$ line, indicating a better out-of-sample performance for the robust estimator.}
    \label{fig:cv_nutrimouse}
\end{figure}

\begin{figure}[!ht]
\centering
    \begin{subfigure}{0.49\textwidth}
    \centering
        \includegraphics[width = 1\textwidth]{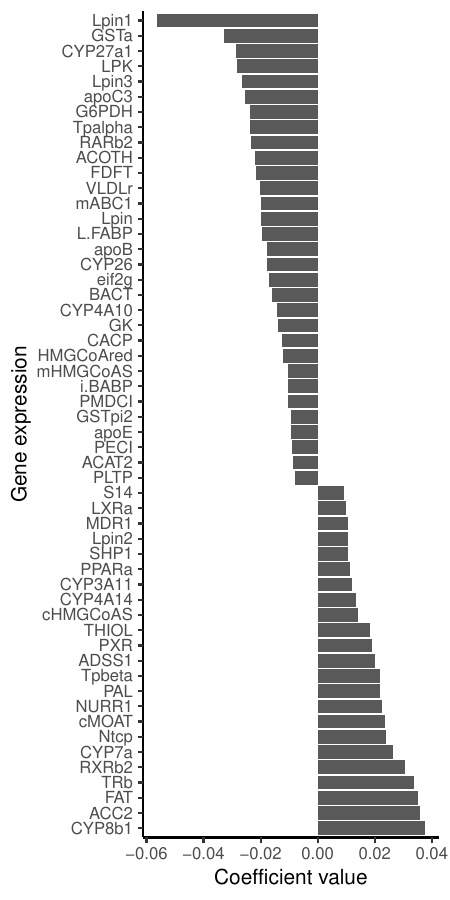}
    \end{subfigure}
    \begin{subfigure}{0.49\textwidth}
    \centering
        \includegraphics[width = 1\textwidth]{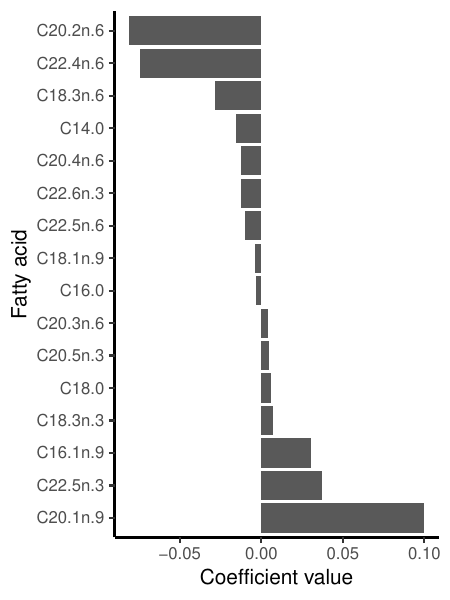}
    \end{subfigure}
    \caption{Estimated linear combinations using the OGK estimator of the covariance matrix for the \texttt{nutrimouse} dataset.}
    \label{fig:nutrimouse}
\end{figure}

The results are given in Table~\ref{tab:nutrimouse}: The residual score for the robust method is much lower, the same holds for the trimmed version. The estimated association is $0.76$ for PMD and $0.89$ for our method using the OGK estimator. Additionally, the contributions to the residual scores for both methods are presented as a scatterplot in Figure~\ref{fig:cv_nutrimouse}. Most of the points are above the 45$^{\circ}$ line, i.e. smaller for the robust method. Similar to \citet{Wilms2015b} we can conclude that the out-of-sample performance of the robust method is better, and therefore present the estimated linear combinations of this method in Figure~\ref{fig:nutrimouse}. 

\begin{table}[!ht]
\centering
\caption{Residual scores and association measure based on data splitting for the \texttt{nutrimouse} dataset.}
\label{tab:nutrimouse}
\begin{tabular}{lccc}\toprule
\textbf{Method} & \textbf{r} & \textbf{r (trimmed)}  & \textbf{Association}\\
\midrule
sparse CCA & $2.15$ & $1.10$ & 0.76\\
OGK & $0.49$ & $0.18$  & 0.89\\
 \bottomrule
\end{tabular}
\end{table}

Out of the $p=120$ gene expressions, $54$ are selected by the algorithm, and out of $q=21$ fatty acids, $16$ are selected. Upon comparison with the selected variables presented by \citet{Wilms2015b}, we can identify the following sets of fatty acids that have been selected by both the SRAR method and ours: C.20.1n.9, C.20.2n.6 and C22.4n.6 with the highest (absolute) coefficients. Among the fatty acids related to the diets of the mice, namely C22:6n-3, C22:5n-3, C22:5n-6, C22:4n-3, and C20:5n-3 \citep[see][]{Martin2007}, 5 are selected by our method. 
For the gene expressions, the coefficient values differ more than for the fatty acids. CYP3A11, which has been found to have a significant influence by \citet{Martin2007}, is selected by both methods. The influence of diet on Lpin1 is also discussed by \citet{Martin2007} and has the highest absolute coefficient in our model.

\subsection{Application in tribology}
Fourier-transform infrared spectroscopy (FTIR) spectra to monitor a lubricant's degradation process and their relation to different indicators for oil condition have been studied by several authors \citep[see, for example,][]{Pfeiffer2022}. However, the goal is to also understand the association between lubricant chemistry and lubrication performance. \citet{Pfeiffer2023} demonstrated how features from optical data of wear scar areas can be extracted and used in a robust partial least-squares (PLS) model to relate the wear scar to oil condition, measured by alteration duration. 

We demonstrate that by using the method presented in this paper, the two high-dimensional datasets can be associated directly. The dataset consists of $n=214$ observations, FTIR spectra with $p=1668$ variables, and HoG (histogram of gradients) feature vectors, with $q=1836$ variables, representing the wear scar images. As previous studies have shown, while sparsity is beneficial for the evaluation of FTIR spectra, it does not yield good results for HoG image features \citep{Pfeiffer2023}. This example also illustrates the flexibility of our approach, being able to choose a sparsity-inducing penalty on one side, while applying $L_2$-regularization on the other. Before proceeding with the analysis (independent from which estimator of the covariance matrix was used), the HoG feature vectors were robustly centered and scaled using median and MAD, respectively. 

We compute the residual score, see Equation~(\ref{eq:residscore}), for a 90\%--10\% split, repeated 5 times.
Here we run the algorithm using the Pearson correlation (sample covariance matrix), and as a robust counterpart we use the OGK estimator of the covariance matrix. 
The resulting residual scores are given in 
Table~\ref{tab:tribology}. It can be observed that the overall out-of-sample performance for the robust estimator is clearly better than for the classical one, while the estimated association is comparable. 

\begin{table}[ht]
\centering
\caption{Residual scores and association measure based on data splitting for 
the tribology dataset.}
\label{tab:tribology}
\begin{tabular}{lccc}\toprule
\textbf{Method} & \textbf{r} & \textbf{r (trimmed)} & \textbf{Association} \\
\midrule
Pearson & $138.51$ & $121.14$ & 0.24\\
OGK & $52.7$ & $48.40$ &  0.21\\
 \bottomrule
\end{tabular}
\end{table}

\begin{figure}[!ht]
\centering
    \begin{subfigure}{0.49\textwidth}
    \centering
        \includegraphics[width = 1\textwidth]{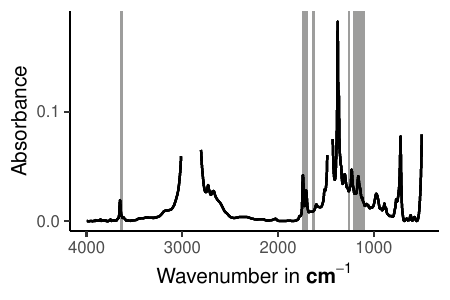}
        \caption{Selected wavenumbers using the sample covariance. }
        \label{fig:FTIR_P}
    \end{subfigure}
        \begin{subfigure}{0.49\textwidth}
    \centering
        \includegraphics[width = 1\textwidth]{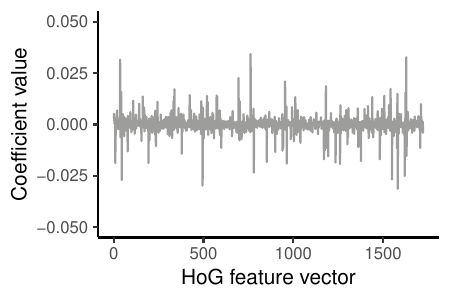}
        \caption{Coefficient values for HoG features using the sample covariance.}
        \label{fig:HOG_P}
    \end{subfigure}
    
    \begin{subfigure}{0.49\textwidth}
    \centering
        \includegraphics[width = 1\textwidth]{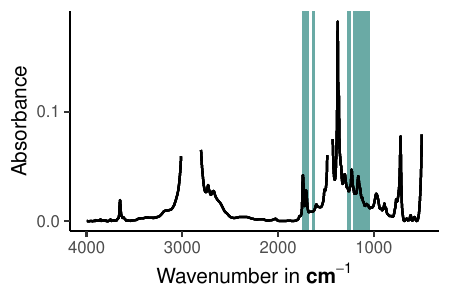}
        \caption{Selected wavenumbers using the OGK estimator.}
        \label{fig:FTIR_OGK}
    \end{subfigure}
    \begin{subfigure}{0.49\textwidth}
    \centering
        \includegraphics[width = 1\textwidth]{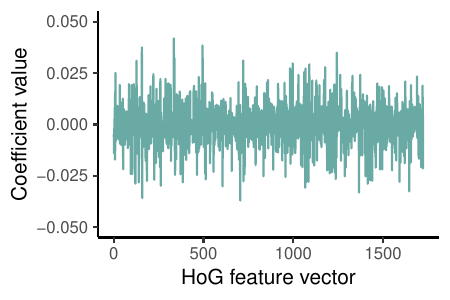}
        \caption{Coefficient values for HoG features using the OGK estimator.}
        \label{fig:HOG_OGK}
    \end{subfigure}
    \caption{Selected wavenumbers (left) using the sample covariance (top, grey) and the OGK estimator (bottom, green). The plots on the right-hand side show the coefficient vector for the HoG features (classical on top, robust at the bottom). For these, no sparsity penalty was included.}
    \label{fig:tribology}
\end{figure}

\begin{figure}[!ht]
\centering
        \includegraphics[width = 0.95\textwidth]{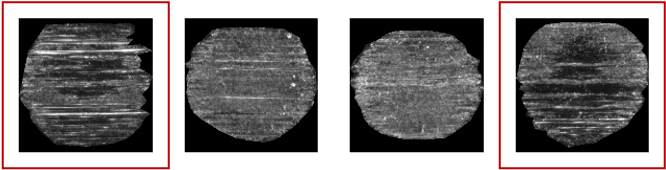}
    \caption{Wear scar areas on the ball for a duration of 552 hours of oil alteration. The framed images correspond to HoG features that were identified as outliers by the OGK estimator.}
    \label{fig:tribology_img}
\end{figure}

In Figure~\ref{fig:tribology} (left), the wavenumbers of the FTIR spectra selected by the non-robust and the robust procedures are displayed. For the non-robust method, 79 wavenumbers are selected, and for the robust one 73, with an overlap of 52 non-zero elements in the two linear combinations.
The selected wavenumbers between 1860-1660 cm$^{-1}$ are known to be related to oxidation processes, while wavenumbers between 3651-3649 cm$^{-1}$ correspond to phenolic antioxidants \citep{Ronai2021}. The selected wavenumbers are similar for the non-robust and robust estimators. For the HoG feature vectors, however, a difference in the size of the coefficients can be observed. The coefficient values for the sample covariance are lower, which can be explained by a stronger regularization parameter being chosen during hyperparameter optimization. Note that the coefficients shown in Figure~\ref{fig:HOG_P} and Figure~\ref{fig:HOG_OGK} are normalized such that $\bm b' \bm C \bm b = 1$. 
As the HoG features have been extracted from images of wear scar areas, outliers can be expected to be present 
\citep[cf.][]{Pfeiffer2023}. While such outliers drive the non-robust method towards over-regularization of these features, the robust method based on the OGK estimator reduces the influence of outliers and is able to determine appropriate coefficient values. An example of wear scar images corresponding to outliers identified by the OGK estimator is given in Figure~\ref{fig:tribology_img}.

\section{Summary and conclusions}

We focused on the problem of maximizing the association between linear combinations of two sets of random variables (two multivariate data sets referring to the same observations). If the association measure is taken as the Pearson correlation, this corresponds to the framework of canonical correlation analysis. However, more general association measures can be considered, even allowing for an identification of monotone rather than just linear relationships. The stated problem can also be formulated as a constrained maximization problem, where the joint covariance of the two random variables is involved, and the coefficients for the linear combinations need to be identified. This is the problem considered in this paper, with two extensions: (i) we aim for robustness of the association measure against outliers in one or both data sets, and (ii) we want to have sparsity in one or both vectors of the linear combinations for applications in high dimensions. 
In this approach, robustness can easily be achieved by plugging in a robustly estimated covariance matrix. Several proposals for this purpose exist in the literature, also for high-dimensional data. We also investigated pairwise estimators of the covariance matrix, e.g.,~based on Spearman's rank correlation.

The main contribution of the paper is the development of an efficient algorithm to solve the optimization problem. The formulation in terms of a Lagrange problem makes the use of a gradient descent algorithm possible, where constraints for the linear combinations can easily be incorporated. The minimum requirements for the penalty functions are that they are convex and that (sub)gradients exist. Other than that, the presented algorithm allows flexibility in the choice of penalty functions, and---as seen in one of the examples---also enables to induce sparsity in only one component, while the other is $L_2$ regularized. 

From a computational perspective, the main advantage of our approach is the scalability in the number of variables. 
Here, we allow both $p$ and $q$ to be larger than the sample size $n$. If $n$ is very small compared to the dimensionality, the quality of the covariance matrix estimation will suffer, leading to unstable results. Furthermore, storage issues may arise if $p$ and/or $q$ are very large (e.g., tens or hundreds of thousands). These issues could potentially be addressed with regularized estimators of the covariance matrix that set certain elements to zero---implemented using sparse matrix data structures---and may therefore be essentially related to implementation details of the algorithm. In any case, the computational cost with increasing dimensionality is essentially driven by the covariance matrix estimation.
In comparison to solving a regression problem repeatedly \citep{Wilms2015b}, or to scanning a larger and larger search space in circles \citep{Alfons2016b}, an algorithm based on gradient descent is more efficient, as only the gradient information needs to be stored. Although the estimation of the plug-in covariance matrix can be computationally demanding, covariance estimation just needs to be done once. This is different from approaches, e.g., based on projection pursuit, where pairwise correlations or association measures need to be computed for all considered projection directions~\citep{Alfons2016b}.

We provided numerical results for the precision and theoretical considerations concerning the existence of a solution of the optimization problem. For this, positive definiteness of the joint covariance matrix is a requirement, however, from the simulations we could see that even when this assumption is violated (high-dimensional setting with pairwise estimators), our algorithm is able to produce comparable or even better results than the alternative techniques. The results emphasize how important a good (and robust!) estimator of the covariance is. As the performance regarding robustness and computation time is dependent on the estimator of the covariance matrix, a sparse and robust estimator could lead to improvements. \citet{Avella2018} show that thresholding methods \citep[see, e.g.,][]{Bickel2008}  have desirable properties when a robust initial estimator is used. Extending the available implementation to exploit the sparsity structure of the covariance will be explored in our upcoming research.

Examples with high-dimensional data sets from biology and tribology underline the usefulness of our approach: It offers flexibility concerning penalty functions depending on the desired sparsity in each of the data sets, desirable robustness properties and maintains manageable computation times. 

The combination of robust estimators and modern optimization techniques yields a powerful toolbox for solving several other common problems in statistics. Especially for robust procedures, where robust estimation (e.g., of a covariance matrix) and optimization can be decoupled, the proposed procedure is very promising. Examples of such extensions are robust principal component analysis and robust linear discriminant analysis, which are topics of our future research.

\section*{Computational details}
The proposed procedure is implemented in the R package \texttt{RobSparseMVA}, which is available from \url{https://github.com/piapfeiffer/RobSparseMVA}. Replication files are publicly available at \url{https://github.com/piapfeiffer/RobSparseMVA-supplement}.

\bibliographystyle{unsrtnat}

\clearpage
\appendix

\section{Preliminaries for optimization} \label{app:prelim}
\begin{definition}[Subgradient]
    For a convex function $f: \mathbb{R}^p \rightarrow \mathbb{R}$, the \emph{subgradient} at $\bm x \in \bm{dom} f$ is defined as the set of vectors $\bm g \in \mathbb{R}^p$, such that for all $\bm z \in \bm{dom} f$, it holds: $f(\bm z)\geq f(\bm x) + \bm g' (\bm z-\bm x)$. If $f$ is convex and differentiable, $\bm g = \nabla f$.
\end{definition}

\noindent
In the following, we use the notation $\partial f$ when referring to the subgradient of a given function $f$.

\begin{definition}[MFCQ constraint qualification]
    The \emph{Mangasarian-Fromovitz constraint qualification (MFCQ)} holds at a minimum $\bm x^*$, if the gradients of the equality conditions $\nabla H(\bm x^*)$ are linearly independent at $\bm x^*$ and there exists a vector $\bm d$, such that $\nabla G(\bm x^*)'\bm d < 0$ and $\nabla H(\bm x^*)'\bm d = 0$ \citep{Mangasarian1967, Ruszczynski2006}. 
\end{definition}

\begin{definition}[KKT conditions]
    Assume that there exists a local minimum $(\bm a^*, \bm b^*)$ of \eqref{eq:CCA_opt}--\eqref{eq:cancor_const_b} and that $F$, $G$, and $G$ are subdifferentiable at $(\bm a^*, \bm b^*)$. Further assume that $(\bm a^*, \bm b^*)$ is a regular point. Then, the Karush-Kuhn-Tucker (KKT) conditions hold \citep[see, e.g.,][]{Boyd2005}:
\begin{itemize}
    \item Stationarity: $\bm 0 \in -\partial F(\bm a^*, \bm b^*) + \partial G(\bm a^*, \bm b^*)'\bs{\mu} + \partial H(\bm a^*, \bm b^*)'\bs{\lambda}$,
    \item Primal feasibility: $G(\bm a^*, \bm b^*) \leq \bm 0$ and $H(\bm a^*, \bm b^*) = 0$,
    \item Dual feasibility: $\bs{\mu} \geq \bm 0$,
    \item Complementary slackness: $\bs{\mu}' G(\bm a^*, \bm b^*) =  0$.
\end{itemize}
\end{definition}

\section{Lemmas}  \label{app:lemmas}
\begin{lemma} \label{lemma:global_min}
    Let $\Omega \subset \mathbb{R}^p \times \mathbb{R}^q$ denote the set of points that satisfy the constraints (\ref{eq:cancor_const_a})--(\ref{eq:cancor_const_penb}) and let $\omega_{\min}(\bm C)$ denote the smallest eigenvalue of matrix $\bm C$. For $c_{a_k} \geq \alpha_{a_k} \sqrt{\nicefrac{p}{\omega_{\min}(\bf C_{xx})}} + (1-\alpha_{a_k}) \nicefrac{1}{\omega_{\min}(\bf C_{xx})}$ and $c_{b_k} \geq \alpha_{b_k} \sqrt{\nicefrac{q}{\omega_{\min}(\bf C_{yy})}} + (1-\alpha_{b_k}) \nicefrac{1}{\omega_{\min}(\bf C_{yy})}$, $\Omega \neq \emptyset$, and the optimization problem (\ref{eq:CCA_opt})--(\ref{eq:cancor_const_b}) attains a global minimum over $\Omega$. 
\end{lemma}
\begin{proof}[Proof of Lemma \ref{lemma:global_min}]
    To show that $\Omega \neq \emptyset$, first note that $\bm C_{xx}$ and $\bm C_{yy}$ are positive definite matrices that induce a norm on $\mathbb{R}^p$ and $\mathbb{R}^q$, respectively. Constraints (\ref{eq:cancor_const_a}) and (\ref{eq:cancor_const_lowa}) are fulfilled by any basis of $\mathbb{R}^p$ that is orthonormal with respect to the norm induced by $\bm C_{xx}$. The same argument can be applied to constraints (\ref{eq:cancor_const_b}) and (\ref{eq:cancor_const_lowb}) with the norm induced by $\bm C_{yy}$. 
Using the inequalities $\| \bm a_k\|_2 \leq \| \bm a_k\|_1 \leq \sqrt{p}\| \bm a_k\|_2$ and $\omega_{\min} \| \bm a_k \|_2^2 \leq \|\bm a_k \|_{\bm C_{xx}}^2$ we can conclude that $\alpha_{a_k} \| \bm a_k\|_ + (1-\alpha_{a_k}) \| \bm a_k\|_2 \leq \alpha_{a_k} \sqrt{\nicefrac{p}{\omega_{\min}(\bm C_{xx})}} + (1-\alpha_{a_k} \nicefrac{1}{\omega_{\min}(\bm C_{xx})})$ and $\nicefrac{1}{c_{a_k}} P_{a_k}(\bm a_k) \leq 1 = \| \bm a_k \|_{\bm C_{xx}}$ for $c_{a_k} \geq \alpha_{a_k} \sqrt{\nicefrac{p}{\omega_{\min}(\bf C_{xx})}} + (1-\alpha_{a_k}) \nicefrac{1}{\omega_{\min}(\bf C_{xx})}$. The same reasoning can be applied to $\bm b_k$ and $\bm C_{yy}$. 
The function $F$ in (\ref{eq:CCA_opt}) is continuous and the feasible region $\Omega \subset \mathbb{R}^p \times \mathbb{R}^q$ is non-empty and compact. Then by Weierstrass' theorem, the function $F$ attains a global minimum over $\Omega$.
\end{proof}

\begin{lemma} \label{lemma:MFCQ}
Let $P_1: \mathbb{R}^p \rightarrow \mathbb{R}$ and $P_2: \mathbb{R}^q \rightarrow \mathbb{R}$ be convex, differentiable functions and let $P_1(\bm a)\geq 0$ for all $\bm a \in \mathbb{R}^p$, $P_2(\bm b)\geq 0$ for all $\bm b \in \mathbb{R}^q$ and $P_1(\bm 0) = P_2(\bm 0) = 0$. The constraints of the optimization problem \eqref{eq:CCA_opt}--\eqref{eq:cancor_const_b} fulfill the Mangasarian-Fromovitz constraint qualification
 (MFCQ) at $(\bm a^*, \bm b^*)$ for all $(\bm a, \bm b) \in \mathbb{R}^p\times \mathbb{R}^q$. 
\end{lemma}
\begin{proof}[Proof of Lemma \ref{lemma:MFCQ}]
    We combine the vector pair $(\bm a, \bm b)$ into a partitioned vector $\bm z = (\bm a, \bm b)' \in \mathbb{R}^{p+q}$. Consider the following partitioning of the identity matrix,
    \begin{align}
    \bm I_{p + q} =  
        \begin{bmatrix}
            \bm I_p & \bm 0 \\
            \bm 0 & \bm I_q
        \end{bmatrix}
        = [\bm E_a | \bm E_b], \label{eq:embedding}
    \end{align}
    which can be used to extract the components of $\bm z$: $\bm a = \bm E_a'\bm z$ and $\bm b = \bm E_b'\bm z$. The gradients of the equality conditions then become
    \begin{align}
        \nabla_z H_1 = 
        \begin{bmatrix}
            \bm C_{xx} & \bm 0 \\
            \bm 0 & \bm 0
        \end{bmatrix}
        \begin{bmatrix}
            \bm a_{1:(k-1)} \\
            \bm 0 
        \end{bmatrix}
        \text{~and~}
        \nabla_z H_2 = 
        \begin{bmatrix}
            \bm 0 & \bm 0 \\
            \bm 0 & \bm C_{yy}
        \end{bmatrix}
        \begin{bmatrix}
            \bm 0 \\
            \bm b_{1:(k-1)}
        \end{bmatrix},
    \end{align}
    and  $s_1 \nabla_z H_1 + s_2 \nabla_z H_2 = \bm 0$ only if $s_1 = s_2 = 0$.

    To see the second part of the condition, let $\bm d = (- \bm a^*, -\bm b^*)'$. Due to the convexity of $P_1$ and $P_2$ and using $P_1(\bm 0) = P_2(\bm 0) = 0$, there holds
    \begin{align}
        -\bm a^* \nabla_a P_1(\bm a^*) - \bm a^* \bm C_{xx} \bm a^* &< 0, \\
        \text{~and~} -\bm b^* \nabla_b P_2(\bm b^*) - \bm b^* \bm C_{yy} \bm b^* &< 0,
    \end{align}
    as well as 
    \begin{align}
         - \bm a^* \bm C_{xx} \bm a_{1:(k-1)} &= 0, \\
        \text{~and~} - \bm b^* \bm C_{yy} \bm b_{1:(k-1)} &= 0,
    \end{align}
    which concludes the proof.
\end{proof}

\section{Algorithms}\label{app:alg}
\begin{algorithm}[h!]
	\caption{Sparse and robust maximum association}\label{alg:robsparse_cca}
	\begin{algorithmic}[1]
        \State Estimate covariance matrices $\bm C_{xx}, \bm C_{yy}, \bm C_{xy}$ \label{alg:estimate_cov}
        \State Initialize $\bm a_k^0$ and $\bm b_k^0$
        \For {$k=1,2,\ldots, \min(p,q)$}
            \State $\bm \lambda^0 \gets H(\bm a_k^0, \bm b_k^0, \bm a_{1:(k-1)}, \bm b_{1:(k-1)})$
    		\While {$\|\bm \lambda^{t+1} - \bm \lambda^{t}\| > \delta$}
    			\State $(\bm a_k^{t+1}, \bm b_k^{t+1}) \gets \argmin \mathcal{L}_c(\bm a_k^t, \bm b_k^t; \bm \mu^t,  \bm \lambda^t)$ \label{alg:optimize_args}
    			\State $\bm \lambda^{t+1} \gets \bm \lambda^{t} + c H(\bm a_k^{t+1}, \bm b_k^{t+1})$
       \State $\bm \mu^{t+1} \gets \bm \mu^{t} + c G(\bm a_k^{t+1}, \bm b_k^{t+1})$
                
                \If {$0.25 \cdot |H(\bm a_k^t, \bm b_k^t) + G(\bm a_k^t, \bm b_k^t)| < |H(\bm a_k^{t+1}, \bm b_k^{t+1}) + G(\bm a_k^{t+1}, \bm b_k^{t+1})|$} \label{alg:reg_if}
                    \State $c \gets 10 \cdot c$ \label{alg:reg}
                \EndIf
                \State $t \gets t + 1$
       \EndWhile
      \EndFor
	\end{algorithmic} 
\end{algorithm}
\begin{algorithm}[h!]
 \caption{Modified AMSGrad algorithm \citep{Reddi2018} for the minimization in line \ref{alg:optimize_args} of Algorithm \ref{alg:robsparse_cca}. This modification uses the partitioned vector $\bm z = (\bm a, \bm b)'$ and allows for simultaneous updates of $\bm a$ and $\bm b$. It also includes an additional thresholding step to create true sparsity.  The maximum and division in lines~\ref{alg:amsgrad_max} and~\ref{alg:amsgrad_div}, respectively, are executed element-wise, and in the thresholding step in lines \ref{alg:thresh_a} -- \ref{alg:thresh_b}, $a_{k_j}$ and $b_{k_j}$ refer to the components of $\bm a_k$ and $\bm b_k$, respectively.} \label{alg:amsgrad}
 \begin{algorithmic}[1]
 \State Input $\bm z_k^0 = (\bm a_k^0, \bm b_k^0)'$ and $\eta_{1i}, \eta_2, \alpha_i$  
 \State Initialize $\bm m_0 = \bm 0$, $\bm v_0 = \bm 0$, $\hat{\bm v_0} = \bm 0$ 
    \While {$\|\nabla_{z}\mathcal{L}_c\| > \delta$}
    \State $\bm g_i \leftarrow \nabla_{z}\mathcal{L}_c$ 
    \State $\bm m_i \gets \eta_{1i}\bm m_{i-1} + (1-\eta_{1i}) \bm g_i$
    \State $\bm v_i \gets \eta_2 \bm v_{i-1} + (1-\eta_2) \bm g_i ^2$
    \State  $ \hat{\bm v_i} \gets \max(\hat{\bm v}_{i-1}, \bm v_i)$ \label{alg:amsgrad_max}
    \State  $\bm z_k^i \gets \bm z_k^{i-1} - \alpha_i \frac{\bm m_i}{\sqrt{\hat{\bm v_i}}}$ \label{alg:amsgrad_div}
    \State $\bm a_k^i \gets \bm E_a' \bm z_k^i$
    \State $\bm b_k^i \gets \bm E_b' \bm z_k^i$
    \State $ d_a^i \gets \frac{\|\bm a_k^i - \bm a_k^{i-1}\|}{\|\bm a_k^{i-1}\|}$ 
    \State $ d_b^i \gets \frac{\|\bm b_k^i - \bm b_k^{i-1}\|}{\|\bm b_k^{i-1}\|}$ 
    \State $i \gets i + 1$
    \EndWhile
    \State $\bar{t}_a \gets \text{avg}[d_a^m]_{m=i}^{i-M+1} + 2\text{sd}[d_a^m]_{m=i}^{i-M+1}$
    \State $\bar{t}_b \gets \text{avg}[d_b^m]_{m=i}^{i-M+1} + 2\text{sd}[d_b^m]_{m=i}^{i-M+1}$
    \State $ \bm a_k \gets [ a_{k_j} \text{~if~} | a_{k_j}| > \bar{t}_a, 0 \text{~otherwise}]_{j=1}^p$ \label{alg:thresh_a}
    \State $ \bm b_k \gets [ b_{k_j}\text{~if~} | b_{k_j}| > \bar{t}_b, 0 \text{~otherwise}]_{j=1}^q$ \label{alg:thresh_b}
    \end{algorithmic}
\end{algorithm}

\clearpage
\section{Complexity} \label{app:complexity}
In Table~\ref{tab:complexity}, the complexities of different approaches are summarized.
For methods that depend on a (robust) covariance estimator, we give both the complexity for the algorithm without covariance estimation and the covariance estimation. The compared methods are PMD, SRAR, and the proposed algorithm, denoted by ``ccaMM'' in combination with different estimators for the covariance matrix For PMD, the first column corresponds to the power iteration to compute the largest eigenvalues, for our proposed method, we give the complexity for gradient descent, for SCCA this corresponds to the complexity of LASSO regression \citep{Tibshirani1996, Efron2004} and for SRAR for sparse LTS regression \citep{Alfons2013, Mount2016}. When it comes to the cost of computing a covariance matrix, we use the sample covariance for PM, SCCA, and the Pearson correlation. The row ``Rank'' corresponds to our method with a covariance matrix based on ranked data (Spearman or Kendall). For the complexity of computing the MCD estimator, we refer to \citet{Bernholt2004}. The OGK estimator is based on the computation of a pairwise robust scale estimator with an additional re-weighting step, which will typically dominate the asymptotic behavior. We assume a scale estimator based on first ordering the data, as suggested by \cite{Maronna2002}, is used.

\begin{table}[!h]
\centering
\caption{Computational complexity of different methods. We denote the number of observations with $n$, and the number of variables with $p$ and $q$, respectively. For ease of notation, we assume $p>q$, which will therefore dominate the asymptotic behavior. The parameter $t$ denotes the number of iterations, and $h$ corresponds to the trimming parameter for LTS. The approximate FastMCD algorithm reduces the complexity to the cost of the repeated computation of a covariance matrix, for $i$ initial subsets and $c$ concentration steps each.} 
\label{tab:complexity}
\begin{tabular}{lcc}
\toprule
\textbf{Method} & \textbf{Complexity} & \textbf{Complexity} \\
 & \textbf{without covariance} & \textbf{for covariance} \\
\midrule
PMD & $O(p^2)$ & $O(np\min(n,p))$ \\
SRAR & $O(\nicefrac{n^p}{h})$ & -  \\
SCCA & $O(p^3) + O(p^2n)$ & $O(np\min(n,p))$ \\
ccaMM - Pearson & $O(tp)$ & $O(np\min(n,p))$  \\
ccaMM - Rank & $O(tp)$ & $O(np\log(n)) + O(np\min(n,p))$  \\
ccaMM - M(R)CD & $O(tp)$ & $O(n^{v + 1})$, $v = \nicefrac{p(p+3)}{2}$  \\
ccaMM - FastM(R)CD & $O(tp)$ &  $O(np\min(n,p)ci)$ \\
ccaMM - OGK & $O(tp)$ & $O(n\log(n)) + O(p^2)$  \\
\bottomrule
\end{tabular}
\end{table}

\section{Bayesian hyperparameter optimization} \label{app:b_opt}
For the basic algorithm, it is assumed that there is a budget of in total $N$ function evaluations. We also need to define a score to be maximized during the hyperparameter optimization, and an appropriate acquisition function. 
 A Gaussian prior is placed on the score function, then its value is observed at $n_0$ points. Until $N$ iterations are reached, the following steps are repeated: (i) update the posterior probability distribution on the score function, (ii) determine the maximum of the acquisition function, and (iii) observe the score value at this parameter configuration.

\begin{figure}[htp]
    \centering
    \includegraphics[width = 0.5\textwidth]{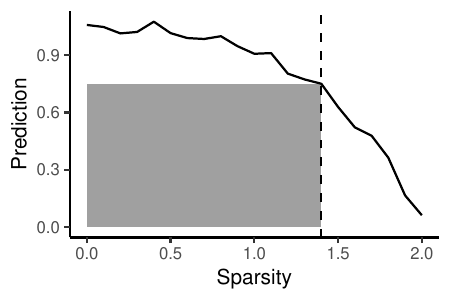}
    \caption{Visualization of the tradeoff product optimization (TPO) criterion (\ref{eq:TPO}) used as a score function in Bayesian hyperparameter optimization. The TPO score corresponds to finding the biggest area under the curve of prediction (robust association measure) over sparsity.}
    \label{fig:TPO}
\end{figure}

We used the implementation in the R package \texttt{ParBayesianOptimization} \citep{R_ParBO} with the expected improvement as acquisition function and the tradeoff product optimization (TPO) as score function. It is similar to the TPO criterion used by \citet{R_pcaPP2024} and models the tradeoff between sparsity in the estimated linear combinations, $\hat{\bm a}_k$ and $\hat{\bm b}_k$, and the estimated value $\hat{\rho}_k$ of the robust association measure. Figure~\ref{fig:TPO} illustrates this criterion.
The original criterion 
\begin{align}
\text{score} = |\hat{\rho}_k| \cdot \left(2 - \frac{\#\{\hat{\bm a}_k\neq 0\}}{p}- \frac{\#\{\hat{\bm b}_k\neq 0\}}{q}\right)
\end{align} 
where $\#\{\hat{\bm a}_k\neq 0\}$ returns the number of non-zero components in $\hat{\bm a}_k$, and similar for $\hat{\bm b}_k$,
can be adapted for non-sparse regularization by including the elastic net parameters $\alpha_{a_k}$ and $\alpha_{b_k}$:  
\begin{align}
\text{score} = |\hat{\rho}_k| \cdot \left(2 - \alpha_{a_k}\frac{\#\{\bm a_k\neq 0\}}{p}- \alpha_{b_k}\frac{\#\{\bm b_k\neq 0\}}{q}\right). \label{eq:TPO}
\end{align} 

Both the sparsity and elastic net parameters can be chosen differently for each $k$ and $\hat{\bm a}_k$ or $\hat{\bm b}_k$, respectively. In our simulations, presented in Section~\ref{sec:simulation_study}, the chosen elastic net parameters $\alpha_{a_k}$ and $\alpha_{b_k}$ are assumed to be the same for each $k$, while the Bayesian optimization procedure to determine the optimal sparsity parameters is run for each $k$.
Furthermore, Section~\ref{sec:precision} of the simulation study is dedicated to the precision of the algorithm, i.e., the performance of the algorithm when the true covariance matrix and theoretically optimal sparsity parameters are provided.

\end{document}